\documentclass[prd,reprint,onecolumn,letterpaper,superscriptaddress,amsfonts,amsmath,amssymb,nofootinbib,longbibliography]{revtex4-2}

\usepackage{float}

\usepackage{braket}
\usepackage{graphicx}
\usepackage{color}
\usepackage[normalem]{ulem}
\usepackage{booktabs}
\usepackage{MnSymbol}
\usepackage{enumerate}
\usepackage{bbold}
\usepackage{subfigure}
\usepackage{subfiles}
\usepackage{units}
\usepackage[usenames,dvipsnames]{xcolor}
\usepackage{placeins}
\usepackage[T1]{fontenc}
\usepackage{amsmath}
\usepackage{paralist}
\usepackage{psfrag}
\usepackage{setspace,tabularx,fancyhdr}
\usepackage{mathtools}
\usepackage[top=2cm,bottom=2cm,left=2cm,right=2cm]{geometry}
\usepackage{enumitem}
\setlist[itemize]{leftmargin=*}
\usepackage{wrapfig}
\usepackage{type1cm}
\usepackage{yfonts}
\usepackage{relsize}
\usepackage{multirow}
\usepackage{tabularx}
\usepackage{array}
\usepackage{rotating}
\usepackage{bigints}
\PassOptionsToPackage{breaklinks}{hyperref}
\usepackage{xurl} 

\usepackage{hyperref}
\hypersetup{colorlinks=true,linkcolor=MidnightBlue,citecolor=blue,filecolor=green,urlcolor=Violet}

\def\beq{\begin{equation}}
\def\eeq{\end{equation}}
\def\bsp{\begin{split}}
\def\esp{\end{split}}
\def\bea{\begin{eqnarray}}
\def\eea{\end{eqnarray}}

\definecolor{mygreen}{rgb}{0.2, 0.8, 0.2}

\renewcommand{\arraystretch}{1.5} 

\newcommand{\IGNORE}[1]{}

\graphicspath{{Images/}}

\begin{document}

\title{Two Step Localization Method for Electromagnetic Followup of LIGO-Virgo-KAGRA Gravitational-Wave Triggers}

\author{Daniel Skorohod}\email{danielsk444@gmail.com}
\affiliation{Department of Physics, Bar Ilan University, Ramat Gan 5290002, Israel}

\author{Ofek Birnholtz}\email{ofek.birnholtz@biu.ac.il}
\affiliation{Department of Physics, Bar Ilan University, Ramat Gan 5290002, Israel}

\begin{abstract}
Rapid detection and follow-up of electromagnetic (EM) counterparts to gravitational wave (GW) signals from binary neutron star (BNS) mergers are essential for constraining source properties and probing the physics of relativistic transients. Observational strategies for these early EM searches are therefore critical, yet current practice remains suboptimal, motivating improved, coordination-aware approaches. We propose and evaluate the \textit{Two-Step Localization} strategy, a coordinated observational protocol in which one wide-field auxiliary telescope and one narrow-field main telescope monitor the evolving GW sky localization in real time. The auxiliary telescope, by virtue of its large field of view, has a higher probability of detecting early EM emission. Upon registering a candidate signal, it triggers the main telescope to slew to the inferred location for prompt, high-resolution follow-up. We assess the performance of \textit{Two-Step Localization} using large-scale simulations that incorporate dynamic sky-map updates, realistic telescope parameters, and signal-to-noise ratio (SNR)-weighted localization contours. For context, we compare \textit{Two-Step Localization} to two benchmark strategies lacking coordination. Our results demonstrate that \textit{Two-Step Localization} significantly reduces the median detection latency, highlighting the effectiveness of targeted cooperation in the early-time discovery of EM counterparts.
Our results point to the most impactful next step: next-generation faster telescopes that deliver drastically higher slew rates and shorter scan times, reducing the number of required tiles; a deeper, truly wide-field auxiliary improves coverage more than simply adding more telescopes.
\end{abstract}
\maketitle
\section{Introduction}\label{section: Introduction}
Binary neutron star (BNS) and neutron star–black hole (NSBH) mergers are key multi-messenger sources, producing both gravitational waves (GWs) and electromagnetic (EM) counterparts across a broad spectral range. The detection of GW170817 \cite{Abbott2017a} and its kilonova established these systems as sites of heavy-element synthesis via the r-process \cite{Kilonova2017}, and opened new avenues for measuring cosmological parameters \cite{H02017} and testing fundamental physics \cite{GRB2017}. Joint analyses that combine GW and EM information have already demonstrated the power of such multi-messenger data for constraining binary and ejecta properties \cite{Coughlin:2018fis}.
Beyond the GW discovery itself, GW170817 triggered an unprecedented global EM-follow-up campaign, spanning ultraviolet, optical, infrared, X-ray, gamma-ray, and radio observations. These observations provided a detailed, multi-wavelength picture of the merger and its ejecta \cite{DES:2017kbs,Cowperthwaite:2017dyu,Nicholl:2017ahq,Chornock:2017sdf,Margutti:2017cjl,Alexander:2017aly,Shappee:2017zly,Troja:2017nqp,Kilpatrick:2017mhz,Hallinan:2017woc,Kasliwal:2017ngb,Evans:2017mmy,Drout:2017ijr,Goldstein:2017mmi,Pozanenko:2017jrn,Savchenko:2017ffs,Kocevski:2017liw,LIGOScientific:2017ync}, establishing GW170817 as the most comprehensively observed transient in modern astrophysics and demonstrating the scientific power of rapid, well-coordinated EM follow-up.
The EM emission accompanying BNS/NSBH mergers includes prompt gamma-ray bursts, kilonovae powered by radioactive ejecta, and longer-timescale afterglows in X-ray, optical, and radio bands. Early emission—especially in UV and optical—encodes valuable information about the merger remnant, outflow properties, and shock-heated ejecta \cite{Siegel_2016, 2017ApJ...848L..23F}. Precursor EM signals, potentially arising from magnetospheric interactions or tidal disruption, could also aid in rapid localization and characterization \cite{2020ApJ...893L...6M, Tiwari2021}.
However, the timely identification of EM counterparts remains a challenge. During GW170817, the associated short GRB was observed by Fermi/INTEGRAL just $\sim$2 seconds after the merger \cite{Abbott2017a}, whereas the first GW alert was only issued with an automated GCN/LVC Notice $\sim$27 minutes post-merger (13{:}08 UT) \cite{LVC_GCN_NoticeG298048}, followed by the first human-readable GCN Circular at $\sim$40 minutes (13{:}21 UT) \cite{LVC_GCN_Circ21505}. This $\sim$2\,s versus $\sim$27\,min gap reflects the fact that satellite detections are triggered and reported autonomously, while the LVC (LIGO-Virgo Collaboration) alerts in 2017 required low-latency analysis combined with human checks before release \cite{IGWN_UserGuide_Alerts}. The initial sky localization then became available only after $\sim$4.5 hours \cite{45h2017}, limiting opportunities for prompt EM follow-up \cite{Sachdeve2020}.
Although current low-latency pipelines routinely deliver public alerts with median latencies of $\sim 30$\,s, GW170817 was an outlier: A loud non-stationary glitch in LIGO–Livingston required extensive data-quality investigations and human vetting, delaying the first GW alert to $\sim$40 minutes and the first skymap to $\sim$4.5 hours after merger \cite{Pankow:2018qpo}.
This motivates the development of fast localization pipelines (e.g., BAYESTAR \cite{2016PhRvD..93b4013S}) and observational strategies capable of capturing early-time signatures.
In this work, we focus on the problem of reducing latency between GW detection and EM observation. Our approach aims to enable rapid, high-resolution follow-up through a coordinated two-step telescope strategy, with the goal of improving detection rates and enhancing the scientific return of multi-messenger observations.\\
Rapid detection of EM counterparts to GW signals from compact binary mergers is essential for probing merger dynamics, constraining ejecta properties, and informing models of relativistic transients \cite{Abbott2017a}. While BNS mergers are the primary focus, NSBH systems may also yield detectable EM emission under favorable conditions \cite{Kawaguchi2016_ApJ}.
This work proposes and evaluates the \textit{Two-Step Localization} strategy, a coordination method designed to reduce the latency of EM counterpart detection following an external GW trigger. The approach utilizes two telescopes: an auxiliary wide-field instrument (e.g., ULTRASAT, with a $204~\mathrm{deg}^2$ FOV \cite{ultrasat2025}) and a main telescope with narrower FOV but significantly finer spatial resolution (e.g., the Swope Telescope, $7~\mathrm{deg}^2$ FOV, $0.435^{\prime\prime}/\mathrm{pixel}$ \cite{Sachdeve2020,2012ApJ...761...22S}); 
for comparison, ULTRASAT operates at $5.4^{\prime\prime}/\mathrm{pixel}$ . In some configurations, the auxiliary telescope may subsequently reconfigure to a narrower FOV and improved resolution. For instance, the LAST telescope supports both a wide mode ($355~\mathrm{deg}^2$, limiting magnitude $\sim 19.6$) and a narrow mode ($7.4~\mathrm{deg}^2$) \cite{Ofek_2023, rasa11v2}, enabling more precise localization once the transient is detected. Notably, both ULTRASAT and LAST are operated by the Weizmann Institute of Science, which can streamline coordination, calibration, and rapid handoffs between the facilities.
Upon receiving a GW alert from the LIGO \cite{Abbott2009}, 
Virgo \cite{Caron1997}, and KAGRA \cite{KAGRA2019} detectors—whose 
low-latency alert products and early-warning performance during O4 
are characterized in \cite{Chaudhary:2023vec}—the auxiliary telescope 
is slewed to the initial sky localization and begins monitoring in 
wide-field mode.
 This configuration maximizes sky coverage at the expense of resolution, increasing the probability of capturing early EM emission.
Once a candidate is identified, the auxiliary telescope triggers the main telescope to slew to the inferred position for high-resolution follow-up. Early GW parameter estimation (e.g., chirp mass, mass ratio, luminosity distance) informs this decision, ensuring that the follow-up telescope's sensitivity is matched to the expected EM signal \cite{Christensen:2022bxb}.
We assess the performance of \textit{Two-Step Localization} using large-scale simulations incorporating dynamic sky-map updates, realistic telescope parameters, and signal-to-noise ratio (SNR)-weighted localization contours \cite{Duverne:2023joq}. For context, we compare \textit{Two-Step Localization} to alternative strategies lacking full coordination. The results demonstrate that \textit{Two-Step Localization} substantially improves detection latency and enhances early-stage EM follow-up capabilities.
\section{Background}\label{sec:background}
Since GW170817, EM follow-up of GW triggers has evolved into increasingly coordinated, strategy-driven campaigns. Wide networks of optical and high-energy facilities now use algorithmic tiling and time-allocation schemes to maximize the probability of catching fast-fading counterparts within large GW localization regions. Tools such as \texttt{gwemopt}\cite{Coughlin:2018lta,Coughlin:2019qkn} were developed to compare and optimize tiling and scheduling strategies across heterogeneous facilities, and quantify how different algorithms can change counterpart detection efficiency by factors of order unity or more. Building on these ideas, dynamic schedulers for target-of-opportunity campaigns (e.g.\ GRANDMA/GROWTH\cite{Almualla:2020hbs}) incorporate visibility constraints, multi-epoch coverage, and the evolving GW skymap into a unified optimization framework. More recently, platforms such as \texttt{tilepy}\cite{SeglarArroyo:2024tilepy} and the M4OPT toolkit enable cross-observatory coordination and mixed-integer optimization of follow-up plans, allowing mid- and small-FoV facilities, as well as future UV missions such as ULTRASAT \cite{ultrasat2025} and NASA's UVEX \cite{Singer:2025UVEX}, to automatically generate observation sequences that account for each event’s GW sky localization and distance distribution; UVEX is optimized for a deep all-sky UV survey with rapid spectroscopic follow-up and an expected 2030 launch, whereas ULTRASAT, with an expected 2027--2028 launch and an extremely wide (204\,deg$^2$) near-UV field of view, is the natural focus of this work.\\
In parallel, GW data-analysis pipelines have developed genuine early-warning capabilities for compact-binary coalescences. Dedicated low-frequency searches and infrastructure designed for pre-merger alerts now aim to identify BNS signals tens of seconds before coalescence and distribute sky localizations fast enough to repoint narrow-field facilities in time for prompt and very-early EM emission\cite{Sachdev:2020lfd,Magee:2021EW}. Forecast studies suggest that, at design sensitivities, current networks could routinely provide tens of seconds of advance warning with $\lesssim 10$--$100~\mathrm{deg}^2$ localizations for nearby BNS mergers\cite{Nitz:2020abbc10}. Together with ongoing efforts to optimize low-latency localization accuracy\cite{Duverne:2023joq}, these developments provide the context for our proposed rapid two-step strategy, which is designed to exploit both early-warning alerts and improved low-latency skymaps to maximize the probability of securing high-quality early-time EM data.

\begin{figure}[H]
    \centering
    \includegraphics[width=0.65\textwidth]{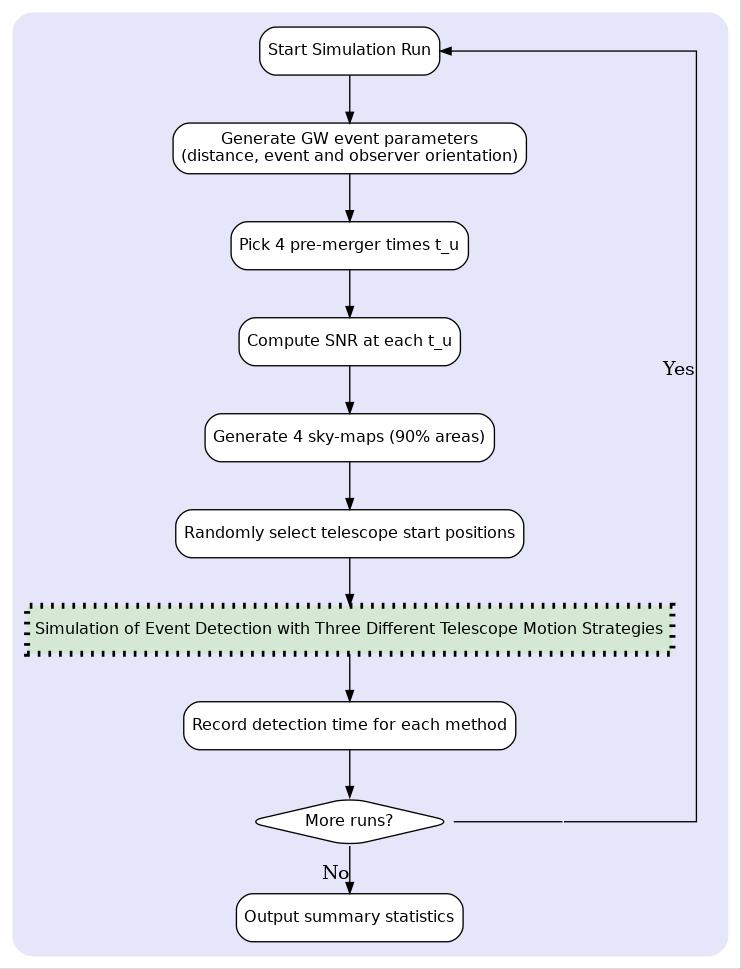}
    \caption{Overall simulation flowchart. Each run proceeds from GW event generation and SNR computation to sky-map updates and detection-time comparison across three coordination strategies. The green, dotted-frame box links to the telescope movement logic in Figure~\ref{fig:tel_movement}.}
    \label{fig:sim_flowchart}
\end{figure}

\section{Method and Approach}\label{Method and Approach}
\begin{figure}[h]
    \centering
    \includegraphics[scale=0.67]{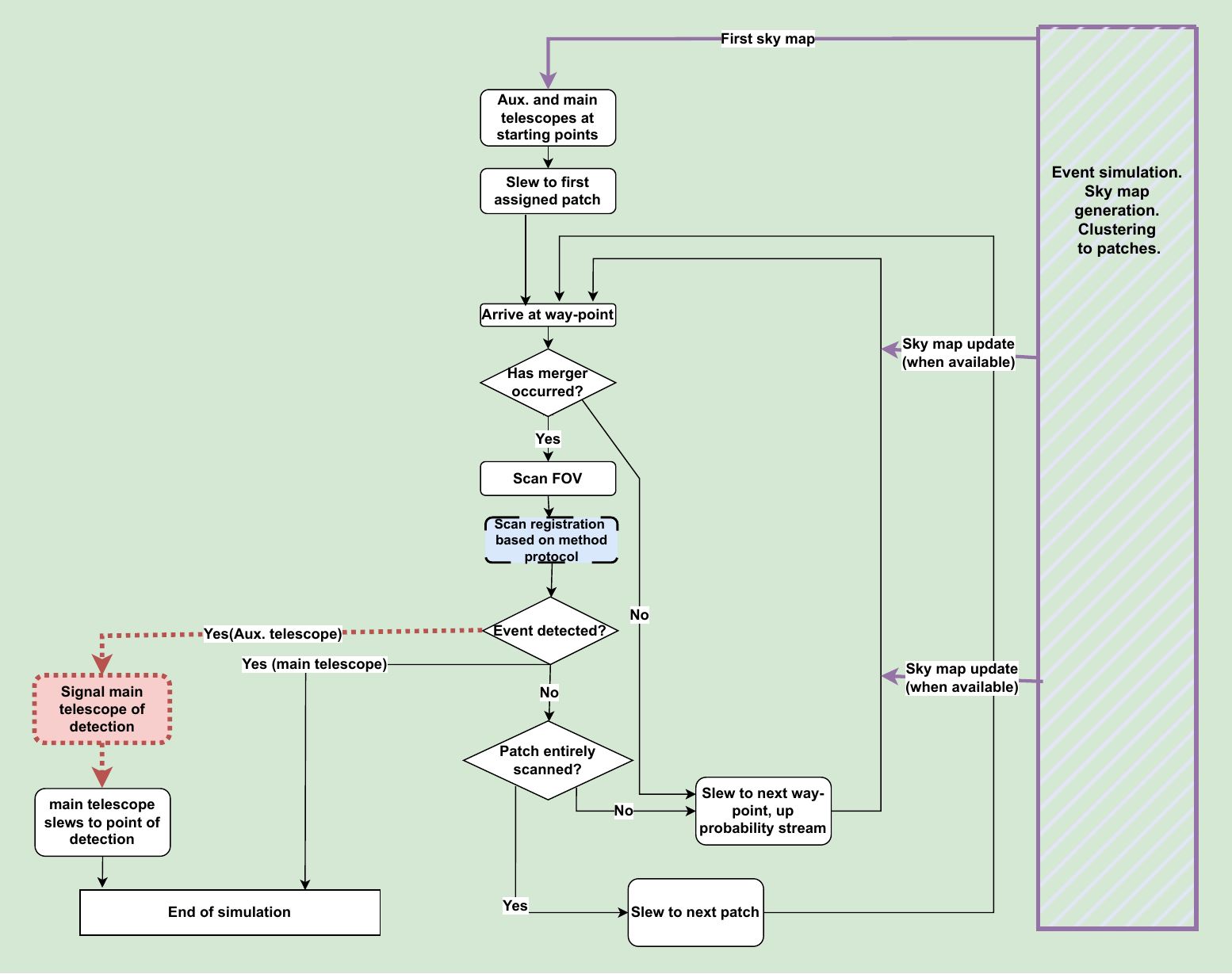}
    \caption{Telescope movement and coordination logic, expanding the green dotted-frame box in Figure~\ref{fig:sim_flowchart}. The purple, dashed-background box connects back to event simulation. The blue dashed-frame box sets communication rules, and the red,dotted, path marks the \textit{Two Step Localization} sequence.}
    \label{fig:tel_movement}
\end{figure}

We evaluate the effectiveness of the proposed \textit{Two Step Localization} method for rapid detection of EM counterparts to BNS mergers. The key question is whether this coordinated telescope movement strategy reduces detection latency compared to uncoordinated approaches. The simulation accounts for pre-merger update times, evolving sky-maps, and SNR evolution, with performance measured by the time to locate the source and the impact of communication protocols between telescopes.
Figures~\ref{fig:sim_flowchart} and~\ref{fig:tel_movement} summarize the simulation architecture. Figure~\ref{fig:sim_flowchart} presents the overall workflow—from GW event generation and SNR calculation to sky-map construction and application of three search strategies. The green, dotted-frame box links directly to the telescope movement logic detailed in Figure~\ref{fig:tel_movement}, while the purple, dashed-background box in Figure~\ref{fig:tel_movement} links back to the event generation stage in Figure~\ref{fig:sim_flowchart}. This cross-referencing highlights the interaction between event simulation and telescope movement modules.
The blue dashed-frame box in Figure~\ref{fig:tel_movement} controls whether telescopes share information on scanned regions (see Sec. \ref{Detection Methods}), and the red, dotted, path marks the \textit{Two Step Localization} sequence—where auxiliary telescope detection immediately redirects the main telescope.
Two LVK observing-scenario analyses help set the broader context for our simulations. Petrov et al.\ \cite{Petrov:2021bqm} use the statistics of O3 public alerts to build simulations of the O4 and O5 observing runs, quantify the resulting distributions of localization areas and other source properties, and examine their implications for electromagnetic follow-up programs such as ZTF, providing a public set of mock localizations for proposal planning. Kiendrebeogo et al.\ \cite{Kiendrebeogo:2023hzf} present updated observing scenarios and multimessenger expectations for the international gravitational-wave network, using event-rate forecasts to relate anticipated GW and counterpart detections to prospective constraints on quantities such as the Hubble constant, r-process enrichment, and the neutron-star equation of state, with the goal of informing future multimessenger campaigns. In addition, Abbott et al.\ \cite{KAGRA:2013rdx} discuss the prospects for observing and localizing gravitational-wave transients with the Advanced LIGO, Advanced Virgo, and KAGRA detectors. In contrast to these network-level, population-based forecasts, our \textit{Two-Step Localization} study follows individual simulated BNS events and models the joint response of only two telescopes—a narrow-field main instrument and a wide-field auxiliary—that could be operated by a single facility or a small collaboration, focusing on how different communication schemes between them affect the time until the EM counterpart is first identified.

\section{Simulation Setup}\label{Simulation}
We simulate pre-merger localizations at four update times ($t_u$) before merger, drawn from 50–60\,s, 40–50\,s, 20–40\,s, and 1–20\,s. SNRs are computed from the PN frequency evolution of the GW signal \cite{2017AnP...52900209A} for $m_1=m_2=1.4\,M_\odot$ \cite{arxiv:2412.05369v1}, using the frequency-domain formalism up to 2PN order \cite{2017EPJP..132...10A}. For BNS and NSBH mergers, the PN expansion remains accurate up to the late inspiral when the mass ratio is modest ($q \lesssim 1$). In this regime, the inspiral dominates the observable signal and tidal disruption effects (TDEs) are well captured without requiring full numerical relativity. Thus, a 2PN description provides sufficiently reliable waveforms for the purposes of pre-merger localization and SNR estimation.\cite{Bernuzzi_2012,Baiotti_2011} 

Detector response functions $F_+$ and $F_\times$ follow \textsc{LALSuite} \cite{LALSuite}, and power spectral densities (PSDs) correspond to LVK O3a \cite{LIGO_T1800044} and to the anticipated O4 \cite{DiCesare:2025wnb} and O5 sensitivities from \cite{LIGO_T2000012}. 
The PSD quantifies detector noise as a function of frequency and is the standard measure of instrumental sensitivity, used to weight the signal contribution with respect to the underlying noise when computing the SNR. 
We retain O3a in our study as an empirically calibrated benchmark tied to the existing LVK BNS and NSBH sample, while O4 and O5 represent progressively improved sensitivities that bracket the performance expected in current and near-future observing runs.
We deliberately test our simulations across different LVK runs to examine the effect of varying noise systematics and the presence of distinct peaks in the noise curves. The more advanced the observing run, the greater the maximum distance reachable for GW detections, owing to improved detector sensitivity. However, the noise curves of these advanced runs are also lower in amplitude, leading to SNRs of comparable magnitude to earlier runs when accounting for the larger distances involved.

For simplicity, we assume that all telescopes used in the simulations are capable of observing events at all distances allowed for each LVK run. In practice, this assumption is optimistic, as the limiting magnitudes of the telescopes are not sufficient to detect events at the farthest distances permitted by the O5 run (325\,Mpc). The Swope telescope has a limiting magnitude of 19.25 for a 15\,s exposure \cite{Coulter:2017wya}, while ULTRASAT reaches a limiting magnitude of 22.4 for a 15\,s exposure \cite{ULTRASAT_IAI_2022} .
The distances adopted here should therefore be viewed as the range over which EM counterparts remain plausibly detectable with the facilities we consider, rather than as hard limits on GW detectability: more distant mergers will occur in O4/O5, but their EM counterparts would in most cases be too faint in the UV/optical for our telescopes and are thus outside the scope of this work.
While the distance to the source can be estimated from the GW early on, along with its direction, and used to estimate within which telescopes' reach it could appear, we live such treatment to future work.

We further assume that the search is scan-time dominated, with a typical scan time of 15\,s per pointing and no additional processing delay. Upon each scan, a detection or non-detection decision is made immediately, and the background sky for every potential transient is assumed to be known in advance. In reality, wide-field time-domain surveys such as ZTF \cite{ZTF} rely on real-time reduction pipelines that perform image subtraction, source association, and candidate vetting, which introduce additional latencies and selection effects. By neglecting these stages we effectively model an idealized reduction chain, so the detection times reported here should be interpreted as optimistic lower bounds relative to current systems. We also assume that telescopes with a narrower (``main'') field of view possess better limiting magnitude and angular resolution than the wide-field (``auxiliary'') telescopes. However, as demonstrated by the Swope and ULTRASAT examples, this is not necessarily the case in reality, where newer wide-field instruments can achieve superior sensitivity despite their larger field of view.
Event distances are drawn from a distribution uniform in comoving volume, with $r_{\rm max}$ set by the chosen LVK run \cite{LIGO_Observing_Plan} (140\,Mpc for O3a, 160\,Mpc for O4, 325\,Mpc for O5). These $r_{\rm max}$ values should not be interpreted as the true GW horizon distances, which extend to substantially larger radii, but as orientation- and sky-location–averaged detection ranges for BNS systems, representative of the typical sensitivity limits quoted in the LVK observing plans. Binary inclination and detector orientation are randomized, affecting both SNR and 90\% sky-map area.
Simulated sky-maps are constructed using past LVK events as morphological templates. For each simulated sky-map update, we first determine the target 90\% credible region area $A_{90}(\mathrm{SNR})$ from the network SNR obtained in our SNR-curve simulation with randomized source parameters (distance, inclination, and sky position). We then randomly select one archived LVK sky-map and order its pixels by posterior probability density. Starting from the highest-probability pixel, we successively accumulate pixels until their total geometric area matches the desired $A_{90}$; the resulting collection of pixels defines the simulated localization region, while all remaining pixels are set to zero probability for simplicity. Finally, we apply a random rotation on the sphere, so that the resulting map is recentered on randomized sky coordinates.
\begin{figure}[ht]
\centering
\includegraphics[width=\linewidth]{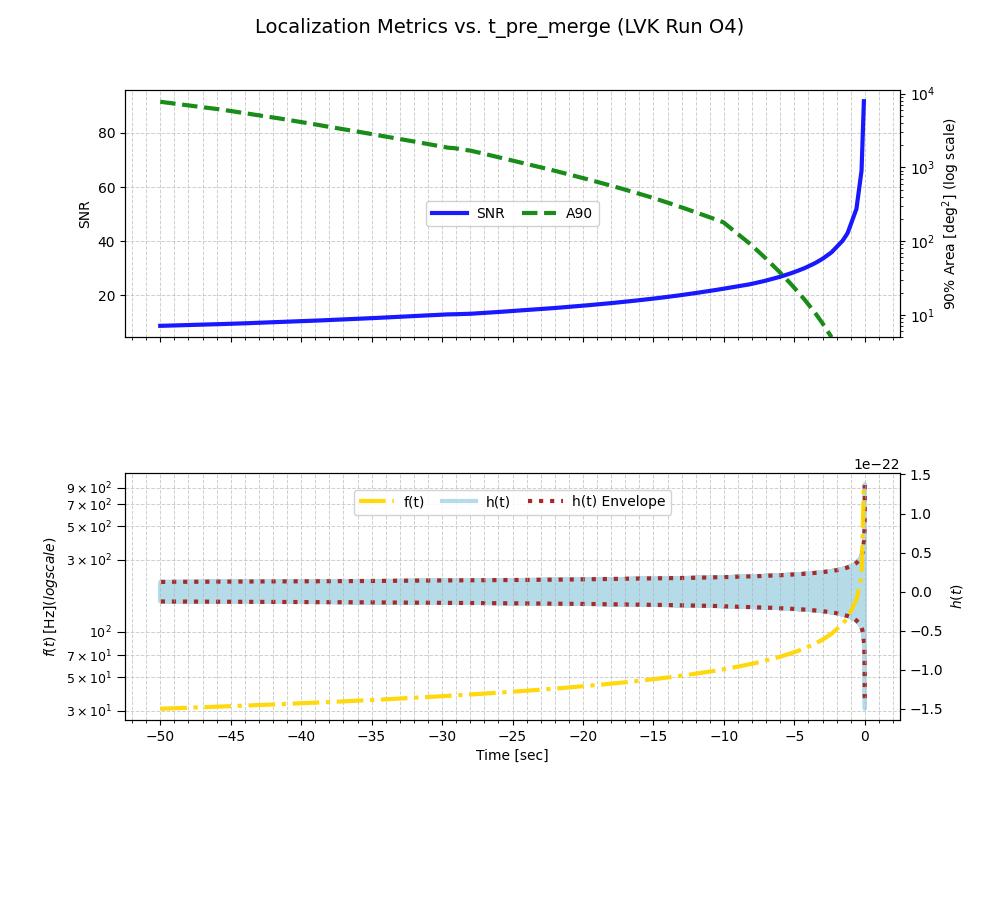}
\caption{\textbf{Localization metrics vs.\ time before merger (LVK O4).} \textbf{Top panel:} SNR (blue, left linear axis) and 90\% credible localization area (green, right logarithmic axis) as functions of the time before merger, \(t_{\rm pre\text{-}merge}\). The SNR rises toward coalescence, while the localization area correspondingly decreases by orders of magnitude as additional in-band cycles accumulate. We truncate the 90\% area curve once it reaches \(A_{90} \simeq 5\,\mathrm{deg}^2\), which is already smaller than the narrowest field of view of the telescopes considered in this work, which is \(A_{90} \simeq 7\,\mathrm{deg}^2\) for Swope telescope. \textbf{Bottom panel:} Gravitational-wave frequency \(f(t)\) versus \(t_{\rm pre\text{-}merge}\). The increasing \(f(t)\) (the GW chirp) drives the SNR growth and thus the improvement—i.e., reduction—of the 90\% area. All curves are generated from our simulations using the O4 noise curve, with SNR computed from the strain model described in this subsection and localization areas obtained from our numerical mapping between SNR, update time \(t_u\), and sky-map size, using the SNR–area fits of \cite{Sachdeve2020}.}
\label{fig:loc_metrics_o5}
\end{figure}

\subsection{Localization metrics vs.\ pre-merger time (O4)}
To visualize how pre-merger information tightens with time, we show one representative possible O4 event:
Figure~\ref{fig:loc_metrics_o5} shows, for a representative BNS event simulated with the LVK O4 noise curve, how the signal frequency \(f(t)\) and the SNR evolve as the system approaches coalescence, and how these trends translate into improved sky localization. As expected from the chirp relation used above, \(f(t)\) increases monotonically as \(|t|\!\to\!0\), and the accumulating in-band cycles yield a steadily rising SNR. Correspondingly, the 90\% credible localization area becomes smaller with time—often by orders of magnitude between tens of seconds pre-merger and the final seconds—because more measurable signal is present where the detectors are most sensitive.

\subsection{Telescope Scanning Model}\label{Telescope Scanning Model}
Waypoints are placed on a grid separated by one main-telescope FOV radius to ensure complete coverage without overlap. Dwell times were set to 15\,s, consistent with ULTRASAT \cite{ultrasat2025} and with the 15\,s single-snap exposure used in Rubin Observatory’s LSST survey (two such exposures per visit, e.g.\ in the $u$ band) \cite{LSST}, while LAST employs 20\,s exposures \cite{Ofek_2023}. The event location is drawn from the final pre-merger sky-map; telescope starting points are random.

At the first $t_u$, each telescope slews to the nearest way-point in its assigned 90\% patch, chosen by weighting detection probability (75\%) and slew time (25\%). If a subsequent $t_u$ occurs before arrival, the path is updated to reflect the reduced search area. Within each patch, telescopes select the nearest neighbor that maximizes total probability in the FOV until the patch is exhausted, then move to the next-highest-probability patch.

\subsection{Detection Methods}\label{Detection Methods}

We compare three strategies, each using a main telescope (Swope-like, $7\,\mathrm{deg}^2$ \cite{Sachdeve2020}) and an auxiliary telescope:  

(1) a wide-field telescope model (“\textit{BIG}”), defined as a hypothetical ZTF-like facility \cite{ZTF}. This auxiliary telescope is simulated with instrumental characteristics of ZTF (cadence, sensitivity, limiting magnitude) but with different field-of-view (FOV) configurations of 100, 200, 400, and 1000\,deg$^2$ to probe the impact of survey depth versus sky coverage. In practice, an existing facility such as LAST \cite{Ofek_2023}, which offers both a wide-FOV mode ($355\,\mathrm{deg}^2$) and a zoomed mode, could serve as a real-world analogue of such a “\textit{BIG}”-class telescope.  

(2) an ULTRASAT-like instrument \cite{ultrasat2025} with $204\,\mathrm{deg}^2$ FOV, representing a space-based UV wide-field surveyor optimized for early detection of transients.  

\begin{table}[H]
\centering
\caption{Summary of detection strategies}
\setlength{\tabcolsep}{5pt}
\renewcommand{\arraystretch}{1.15}
\begin{tabular}{@{} l c c p{0.42\linewidth} @{}}
\hline
Method &
\begin{tabular}[c]{@{}c@{}}Communication\\ on Scan\end{tabular} &
\begin{tabular}[c]{@{}c@{}}Communication\\ on Detection\end{tabular} &
Detection Trigger \\
\hline
Partial & Yes & No  & Main telescope only \\
None    & No  & No  & Main telescope only \\
\textit{Two-Step Localization} & Yes & Yes & Aux.\ triggers Main;\\ & & & ~~Main can detect independently \\
\hline
\end{tabular}
\end{table}


Detection time throughout our study is defined as the elapsed time until the main telescope (Swope-like) identifies the EM counterpart, while the auxiliary telescope contributes by reducing the localization region and guiding the main telescope toward the most probable sky areas.

\paragraph{Partial Communication.} Telescopes avoid re-scanning each other’s fields, but auxiliary detections are not relayed to the main telescope.

\paragraph{No Communication.} Telescopes operate entirely independently, with no sharing of scanned fields or detections.

\paragraph{Two Step Localization.}  
Full communication. If the auxiliary telescope detects the event first, it transmits the coordinates to the main telescope for immediate follow-up. The main telescope, however, retains the ability to detect the event on its own if it happens upon the correct region before receiving a trigger. This dual pathway ensures both rapid response to auxiliary detections and independent discovery capability.

\section{Results}\label{section: Results}
We analyze detection times for 942 simulated BNS mergers (see Sec.~\ref{Simulation}). Each event is evaluated under three strategies—\textit{Partial Communication Method}, \textit{No Communication Method}, and our proposed \textit{Two Step Localization Method}—and five auxiliary configurations: four FOV settings for the ZTF-like “\textit{BIG}” telescope ($100~\text{deg}^2$, $200~\text{deg}^2$, $400~\text{deg}^2$, $1000~\text{deg}^2$) and \textit{ULTRASAT} (see Subsec.~\ref{Detection Methods}). 
\begin{table}[H]
    \centering
    \caption{Summary statistics for the simulated BNS population at the four pre-merger
    update times $t_{\rm pre}^{(i)}$ ($i=1\ldots 4$) for each LVK observing run. 
    Listed are the mean and standard deviation ($\langle\cdot\rangle \pm \sigma$) and 
    the median ($\tilde{\cdot}$) of the SNR and 90\% credible area $A_{90}$ at each update.}
    \label{tab:summary_stats}
    \begin{tabular}{ccccccc}
        \hline
        LVK run &
        Update $i$ &
        $\langle t_{\rm pre}^{(i)} \rangle \pm \sigma_t$ [s] &
        $\langle{\rm SNR}^{(i)}\rangle \pm \sigma_{\rm SNR}$ &
        $\widetilde{{\rm SNR}}^{(i)}$ &
        $\langle A_{90}^{(i)} \rangle \pm \sigma_{A_{90}}$ [deg$^2$] &
        $\widetilde{A}_{90}^{(i)}$ [deg$^2$] \\
        \hline
        O3a &
        1 & $51.44 \pm 0.87$ & $8.57 \pm 17.25$ & $3.29$ &
        $(4.0 \pm 5.7)\times10^{5}$ & $1.0\times10^{5}$ \\
        O3a &
        2 & $41.55 \pm 0.85$ & $10.20 \pm 20.38$ & $3.98$ &
        $(2.5 \pm 3.6)\times10^{5}$ & $5.9\times10^{4}$ \\
        O3a &
        3 & $21.51 \pm 0.86$ & $16.61 \pm 33.28$ & $6.50$ &
        $(5.6 \pm 8.4)\times10^{4}$ & $1.1\times10^{4}$ \\
        O3a &
        4 & $6.47 \pm 0.86$ & $32.43 \pm 65.94$ & $12.50$ &
        $(4.0 \pm 6.8)\times10^{3}$ & $6.1\times10^{2}$ \\
        \hline
        O4  &
        1 & $51.51 \pm 0.87$ & $10.86 \pm 13.85$ & $8.05$ &
        $(1.9 \pm 3.1)\times10^{5}$ & $9.4\times10^{3}$ \\
        O4  &
        2 & $41.51 \pm 0.86$ & $12.90 \pm 16.36$ & $9.68$ &
        $(1.2 \pm 1.9)\times10^{5}$ & $5.1\times10^{3}$ \\
        O4  &
        3 & $21.55 \pm 0.85$ & $19.84 \pm 25.21$ & $14.51$ &
        $(3.0 \pm 5.1)\times10^{4}$ & $1.0\times10^{3}$ \\
        O4  &
        4 & $6.44 \pm 0.88$ & $33.64 \pm 42.68$ & $24.89$ &
        $(3.1 \pm 5.5)\times10^{3}$ & $5.8\times10^{1}$ \\
        \hline
        O5  &
        1 & $51.50 \pm 0.84$ & $5.89 \pm 10.99$ & $2.13$ &
        $(4.8 \pm 6.6)\times10^{5}$ & $3.3\times10^{5}$ \\
        O5  &
        2 & $41.49 \pm 0.88$ & $7.11 \pm 13.13$ & $2.60$ &
        $(2.9 \pm 4.1)\times10^{5}$ & $1.9\times10^{5}$ \\
        O5  &
        3 & $21.53 \pm 0.86$ & $11.73 \pm 21.68$ & $4.15$ &
        $(6.4 \pm 9.6)\times10^{4}$ & $4.1\times10^{4}$ \\
        O5  &
        4 & $6.41 \pm 0.89$ & $22.96 \pm 42.07$ & $8.56$ &
        $(4.3 \pm 7.1)\times10^{3}$ & $1.9\times10^{3}$ \\
        \hline
    \end{tabular}
\end{table}

\subsection{Examples of Telescope Motion for Event Detection}\label{Examples of telescope motion for event detection}

The figures compare the search dynamics for the event’s EM signal under the three strategies. The auxiliary starts at the red hexagon (bold dots show its path); the main starts at the orange hexagon ($\times$ marks). Color along the paths encodes time since the first sky-map update alarm. Red/orange triangles mark the closest initial way-points where scanning begins.

\subsection*{Animated visualization of telescope surveys}

To complement the static figures shown below, we provide an animated
visualization of the telescope trajectories and evolving sky localization
for all three coordination strategies: \textit{Two-Step Localization},
\textit{Partial Communication}, and \textit{No Communication}. The animation
illustrates how the GW sky-map is updated in time and how the auxiliary
wide-field and main narrow-field telescopes respond as their scanned regions
gradually build up. Key moments, such as the merger and subsequent sky-map
updates, are annotated in the video. The animation is available as
Supplementary Video~1 (\texttt{Telescope\_Surveys.gif}) at
\href{https://drive.google.com/file/d/11rGAdwOz0R53FNEzHbYUpG7qAgGuZXEl/view?usp=sharing}
{this link}.

\subsubsection{Simulation 1. - \textit{Two Step Localization Method}}\label{Simulation - Two step method}
\begin{figure}[H]
    \centering
    \includegraphics[scale=0.15]{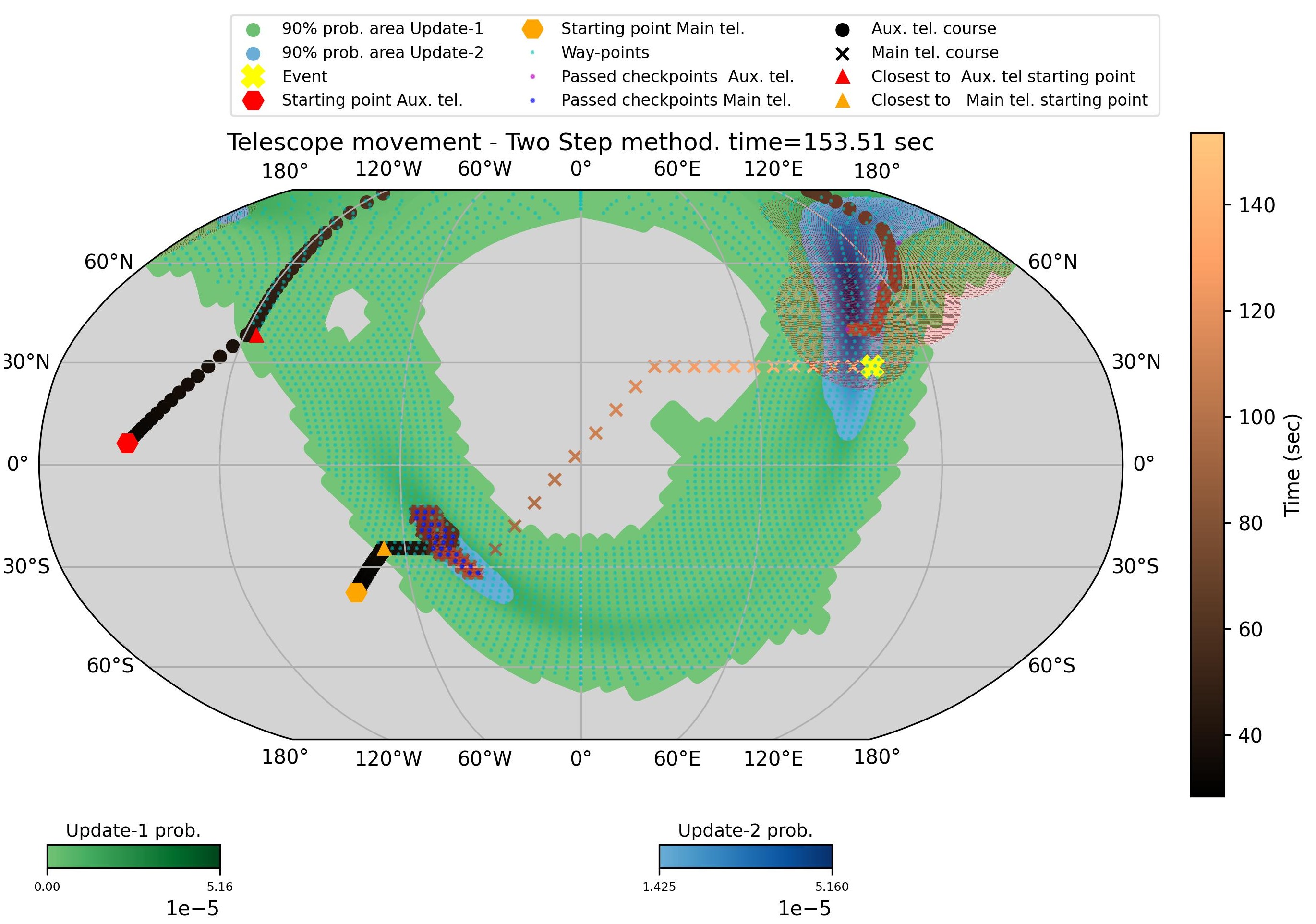}
\caption{Telescope motion and event detection using the \textit{Two Step Localization Method}, with "\textit{BIG}" as the auxiliary telescope configured with a $1000~\text{deg}^2$ FOV. The event is detected $153.5~\text{sec}$ after the first alarm, which was triggered $51.2~\text{sec}$ before the merger—i.e., $102.3~\text{sec}$ after the merger. Localization is progressively refined through subsequent sky-map updates. The third sky-map update—providing the first meaningful localization (see Sec. \ref{Simulation})—marks the start of telescope motion. The time color-bar begins at $28.4~\text{sec}$, representing the time elapsed from the first to the third sky-map update. In this timeline, the merger occurs at $51.2~\text{sec}$.}
    \label{fig:TwoStepMethod1}
\end{figure}

The telescope trajectories and event detection sequence for this method are illustrated in Figure~\ref{fig:TwoStepMethod1}. The full timeline of events is detailed in Table~\ref{tab:dual_motion_timeline}.
The first alarm occurs $51.2~\text{s}$ pre-merger at SNR $3.4$, insufficient for localization (see Sec. \ref{Simulation}); a second at $41.0~\text{s}$ with SNR $4.1$ is likewise uninformative. The third update at $22.8~\text{s}$ (SNR $6.0$, $14403~\text{deg}^2$) initiates motion; the fourth at $6.5~\text{s}$ (SNR $10.5$, $1029~\text{deg}^2$) further refines the search region. The event is detected $153.5~\text{s}$ after the first alarm ($+102.3~\text{s}$ post-merger), as summarized in Figure~\ref{fig:TwoStepMethod1} and Table~\ref{tab:dual_motion_timeline}.

\noindent\textbf{Telescope Motion Timeline — \textit{Two Step Localization Method}:\\}
\noindent After the third update, the auxiliary slews from $(209.06,6.32)$ toward the top-left; the main from $(279.97,-37.45)$ toward the bottom-left. The main reaches its first scan position at $-9.5~\text{s}$, the auxiliary at $-8.2~\text{s}$. Following the fourth update ($-6.5~\text{s}$), both adjust their targets. The main starts scanning at $+3.2~\text{s}$; the auxiliary at $+11.0~\text{s}$ with three $15~\text{s}$ stops (Subsec.~\ref{Telescope Scanning Model}). The auxiliary detects at $(100.54,6.32)$ at $+51.1~\text{s}$, integrates $15~\text{s}$, triggers the main at $+66.1~\text{s}$, and the main confirms at $+102.3~\text{s}$, as summarized in Table~\ref{tab:dual_motion_timeline}.

\begin{table}[H]
\centering
\caption{Simulation 1. - Timeline of Telescope Motion and Detection Events - \textit{Two Step Localization Method}.
$\boldsymbol{t_{\text{Two-Step-Localization-Method}} = t_{\text{No-Communication-Method}} - 254.5~\text{sec}}$}
\label{tab:dual_motion_timeline}
\renewcommand{\arraystretch}{1.2}
\setlength{\tabcolsep}{3pt}
\footnotesize
\begin{tabular}{|p{1.5cm}|p{1.7cm}|p{5.5cm}|p{5.5cm}|}
\hline
\textbf{Since Merger} & \textbf{Since First Alarm} & \textbf{Auxiliary Telescope (\textit{BIG})} & \textbf{Main Telescope} \\
\hline
$-51.2$ s & $0.0$ s & \multicolumn{2}{p{11.2cm}|}{First sky-map update. SNR = 3.4. No meaningful localization.} \\
\hline
$-41.0$ s & $10.2$ s & \multicolumn{2}{p{11.2cm}|}{Second sky-map update. SNR = 4.1. No meaningful localization.} \\
\hline
$-22.8$ s & $28.4$ s & Third sky-map update (green). SNR = 6.0, area = $14403~\text{deg}^2$. Motion begins from $(209.06,6.32)$. & Third sky-map update (green). SNR = 6.0, area = $14403~\text{deg}^2$. Motion begins from $(279.97,-37.45)$. \\
\hline
$-9.5$ s & $41.7$ s & --- & Arrives at first scan position (bottom-left). \\
\hline
$-8.2$ s & $43.0$ s & Arrives at first scan position (top-left). & --- \\
\hline
$-6.5$ s & $44.7$ s & Fourth sky-map update (blue). SNR = 10.5, area = $1029~\text{deg}^2$. Slews toward upper-right of the search area. & Fourth sky-map update (blue). SNR = 10.5, area = $1029~\text{deg}^2$. Slews toward bottom-left of the search area. \\
\hline
$0.0$ s & $51.2$ s & \multicolumn{2}{|c|}{Merger occurs.} \\
\hline
$+3.2$ s & $54.4$ s & --- & Begins scanning updated region. \\
\hline
$+11.0$ s & $62.2$ s & Begins scanning with three $15~\text{s}$ stops (see Subsec. \ref{Telescope Scanning Model}). & --- \\
\hline
$+51.1$ s & $102.3$ s & Detects event at $(100.54,6.32)$ at end of third stop. & --- \\
\hline
$+66.1$ s & $117.3$ s & Sends trigger to main telescope after $15~\text{s}$ integration. & Main telescope receives trigger and slews toward event. \\
\hline
$+102.3$ s & $153.5$ s & --- & Detects event after slewing to the source. \\
\hline
\end{tabular}
\end{table}

\clearpage

\vspace{-20pt}
\subsubsection{Simulation 2. - \textit{Partial Communication Method}}\label{Simulation - Partial Communication Method}
\vspace{-20pt}

\begin{table}[H]
\centering
\caption{Simulation 2. - Timeline of Telescope Motion and Detection Events - \textit{Partial Communication Method}.
$\boldsymbol{t_{\text{Partial-Communication-Method}} = t_{\text{No-Communication-Method}} - 158~\text{sec}}$}
\label{tab:pcm_motion_timeline}
\renewcommand{\arraystretch}{1.2}
\setlength{\tabcolsep}{3pt}
\footnotesize
\begin{tabular}{|p{1.5cm}|p{1.7cm}|p{5.5cm}|p{5.5cm}|}
\hline
\textbf{Since Merger} & \textbf{Since First Alarm} & \textbf{Auxiliary Telescope (\textit{BIG})} & \textbf{Main Telescope} \\
\hline
$-51.2$ s & $0.0$ s & \multicolumn{2}{p{11.2cm}|}{First sky-map update. SNR = 3.4. No meaningful localization.} \\
\hline
$-41.0$ s & $10.2$ s & \multicolumn{2}{p{11.2cm}|}{Second sky-map update. SNR = 4.1. No meaningful localization.} \\
\hline
$-22.8$ s & $28.4$ s & Third sky-map update (green). SNR = 6.0, area = $14403~\text{deg}^2$. Motion begins from $(209.06,6.32)$. & Third sky-map update (green). SNR = 6.0, area = $14403~\text{deg}^2$. Motion begins from $(279.97,-37.45)$. \\
\hline
$-9.5$ s & $41.7$ s & --- & Arrives at first scan position (bottom-left). \\
\hline
$-8.2$ s & $43.0$ s & Arrives at first scan position (top-left). & --- \\
\hline
$-6.5$ s & $44.7$ s & Fourth sky-map update (blue). SNR = 10.5, area = $1029~\text{deg}^2$. Slews toward upper-right of the search area. & Fourth sky-map update (blue). SNR = 10.5, area = $1029~\text{deg}^2$. Slews toward bottom-right of the search area. \\
\hline
$0.0$ s & $51.2$ s & \multicolumn{2}{|c|}{Merger occurs.} \\
\hline
$+3.2$ s & $54.4$ s & --- & Continues scanning bottom-right region. \\
\hline
$+11.0$ s & $62.2$ s & Begins scanning with three $15~\text{s}$ stops (see Subsec. \ref{Telescope Scanning Model}). & --- \\
\hline
$+51.1$ s & $102.3$ s & Detects event at $(100.54,6.32)$ but does not send trigger; halts scanning. & --- \\
\hline
$+100$ s & $151.2$ s & --- & Completes assigned bottom-right search, slews toward top-right updated region. \\
\hline
$+198.8$ s & $250.0$ s & --- & Detects event after scanning second region, including $15~\text{s}$ integration. \\
\hline
\end{tabular}
\end{table}

\begin{figure}[H]
    \centering
    \includegraphics[scale=0.15]{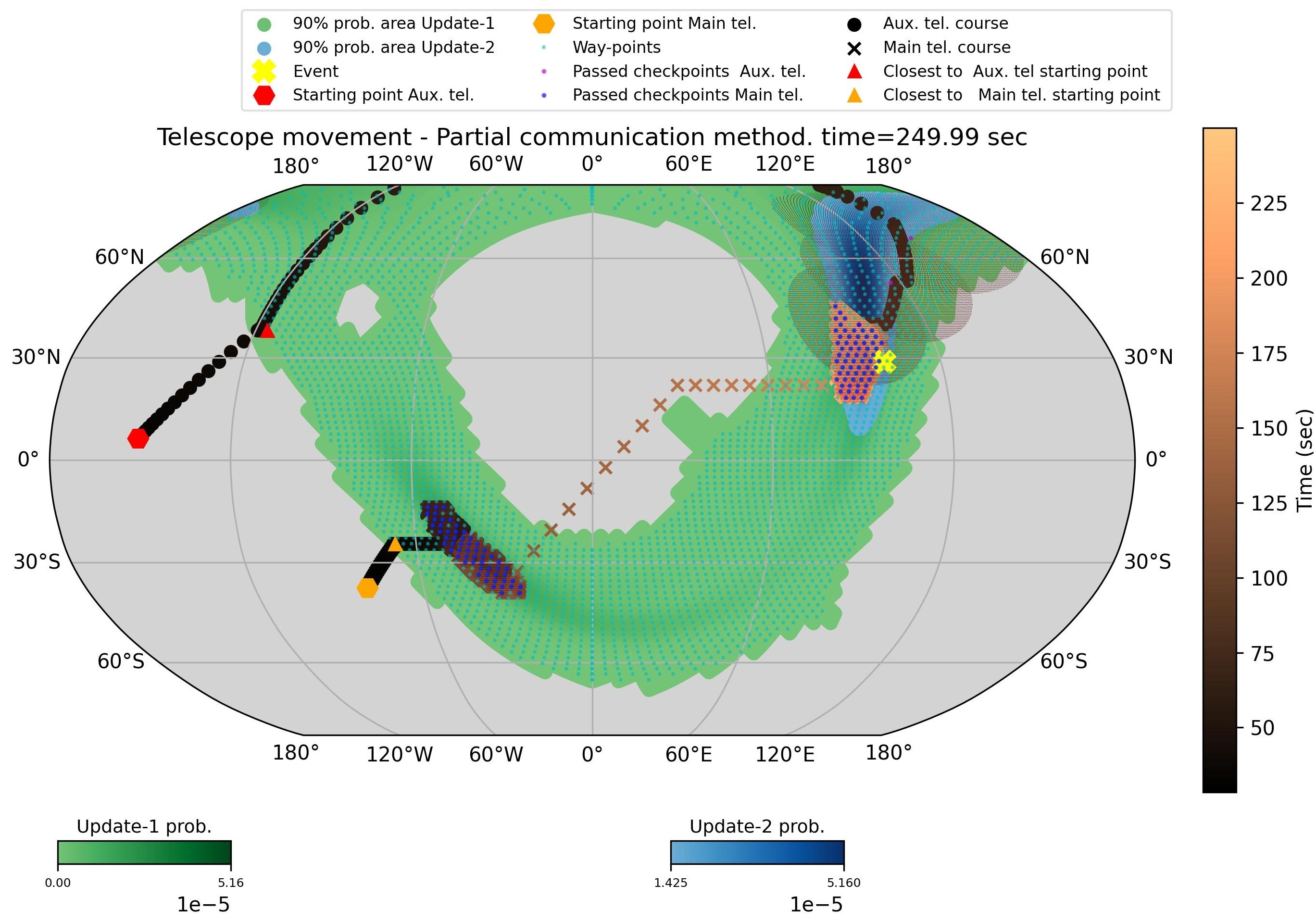}
\caption{Telescope motion and event detection using the \textit{Partial Communication Method}, with "\textit{BIG}" as the auxiliary telescope configured with a $1000~\text{deg}^2$ FOV. The event is detected $250.0~\text{sec}$ after the first alarm, which was triggered $51.2~\text{sec}$ before the merger—i.e., $198.8~\text{sec}$ after the merger. Localization is progressively refined through subsequent sky-map updates. In contrast to the \textit{Two Step Localization Method}, the auxiliary telescope does not signal the main telescope upon detection, resulting in a longer search before the main telescope detects the event.}
     \label{fig:PartialCommunicationMethod1}
\end{figure}

The results of this strategy are shown in Figure~\ref{fig:PartialCommunicationMethod1}, with the detailed sequence of telescope operations summarized in Table~\ref{tab:pcm_motion_timeline}.
The SNR/update sequence is identical to the \textit{Two Step Localization} case; differences arise solely from operations. The auxiliary detects at $(100.54,6.32)$ but does not trigger the main, which must complete its current region and then scan the updated region—prolonging detection to $250.0~\text{s}$ after the first alarm, as seen in Figure~\ref{fig:PartialCommunicationMethod1} and Table~\ref{tab:pcm_motion_timeline}.\\
\noindent\textbf{Telescope Motion Timeline — \textit{Partial Communication Method}:\\}
\noindent Motion begins at the third update ($-22.8~\text{s}$). The main reaches its first scan position at $-9.5~\text{s}$; the auxiliary at $-8.2~\text{s}$. After the fourth update, both redirect. The auxiliary starts scanning at $+11.0~\text{s}$, detects at $+51.1~\text{s}$, but halts without alerting. The main completes its assigned region, slews to the updated sector, and finally detects at $+198.8~\text{s}$ post-merger, consistent with Table~\ref{tab:pcm_motion_timeline}.

\subsubsection{Simulation 3. - \textit{No Communication Method}}\label{Simulation - No Communication Method}

\begin{figure}[H]
    \centering
    \includegraphics[scale=0.15]{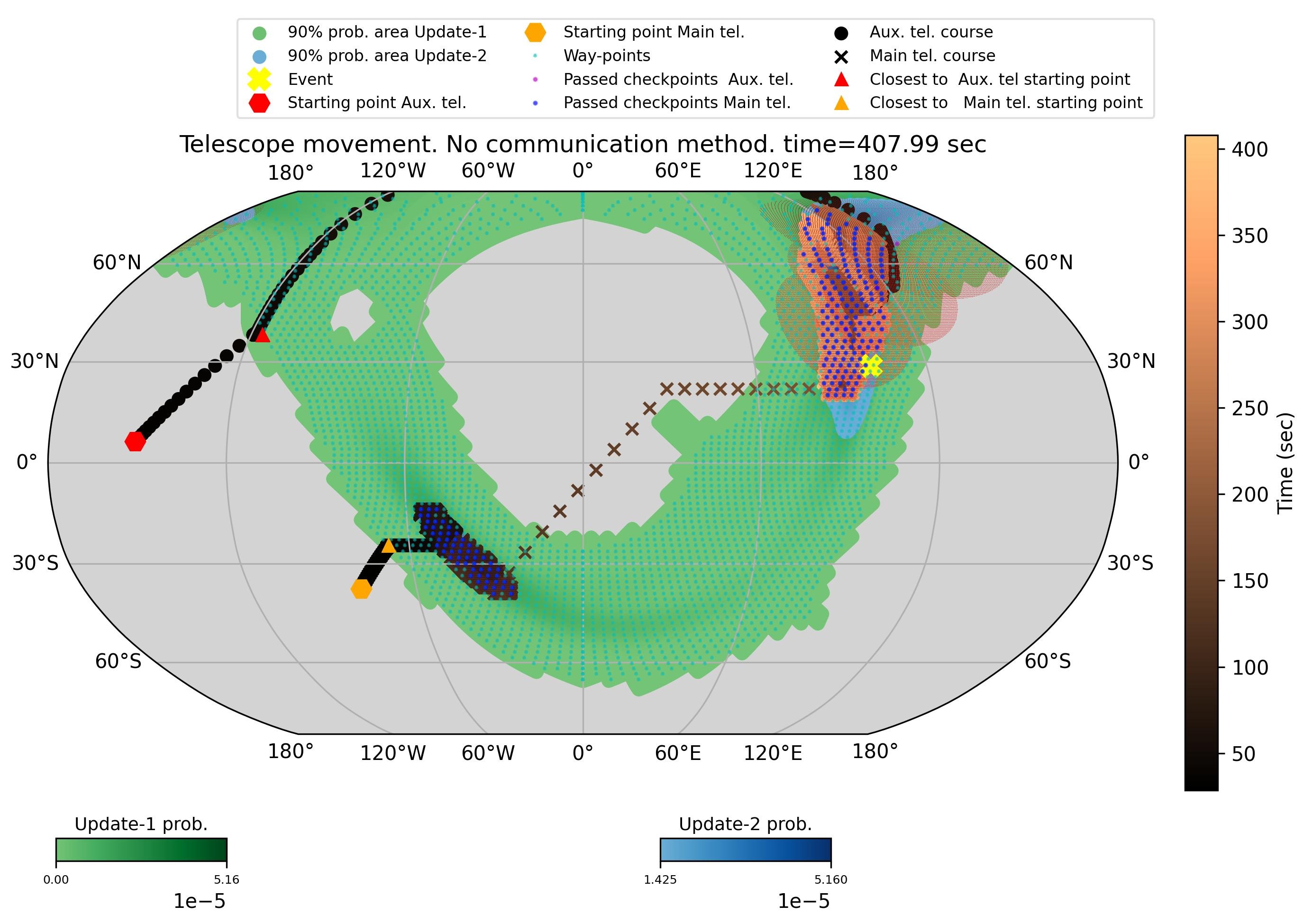}
\caption{Simulation 3. -Telescope motion and event detection using the \textit{No Communication Method}, with "\textit{BIG}" as the auxiliary telescope configured with a $1000~\text{deg}^2$ FOV. The event is detected $408.0~\text{sec}$ after the first alarm, which was triggered $51.2~\text{sec}$ before the merger—i.e., $356.8~\text{sec}$ after the merger. Localization is progressively refined through subsequent sky-map updates. In this method, no information is exchanged between telescopes—neither detection triggers nor negative results—resulting in the longest detection time of the three strategies.}
    \label{fig:NoCommunicationMethod1}
\end{figure}

The operational differences and outcomes of this approach are depicted in Figure~\ref{fig:NoCommunicationMethod1}, while the complete detection timeline is given in Table~\ref{tab:ncm_motion_timeline}.
Here, neither positive detections nor cleared regions are shared. With the same updates/SNRs as above, the auxiliary detects at $+51.1~\text{s}$ but does not communicate; the main completes its initial region and then scans the updated sector without knowledge of cleared areas, yielding the longest delay: $408.0~\text{s}$ after the first alarm, consistent with Figure~\ref{fig:NoCommunicationMethod1} and Table~\ref{tab:ncm_motion_timeline}.

\noindent\textbf{Telescope Motion Timeline — \textit{No Communication Method}:\\}
\noindent Motion starts at the third update. The main arrives at $-9.5~\text{s}$; the auxiliary at $-8.2~\text{s}$. After the fourth update, both redirect. The auxiliary begins scanning at $+11.0~\text{s}$ and detects at $+51.1~\text{s}$ but does not alert. The main, lacking any shared information, finishes its assigned region, slews to the updated sector, and—after an extended pass—detects at $+356.8~\text{s}$ post-merger, as reported in Table~\ref{tab:ncm_motion_timeline}.

\begin{table}[H]
\centering
\caption{Timeline of Telescope Motion and Detection Events – \textit{No Communication Method}}
\label{tab:ncm_motion_timeline}
\renewcommand{\arraystretch}{1.2}
\setlength{\tabcolsep}{3pt}
\footnotesize
\begin{tabular}{|p{1.5cm}|p{1.7cm}|p{5.5cm}|p{5.5cm}|}
\hline
\textbf{Since Merger} & \textbf{Since First Alarm} & \textbf{Auxiliary Telescope (\textit{BIG})} & \textbf{Main Telescope} \\
\hline
$-51.2$ s & $0.0$ s & \multicolumn{2}{p{11.2cm}|}{First sky-map update. SNR = 3.4. No meaningful localization.} \\
\hline
$-41.0$ s & $10.2$ s & \multicolumn{2}{p{11.2cm}|}{Second sky-map update. SNR = 4.1. No meaningful localization.} \\
\hline
$-22.8$ s & $28.4$ s & Third sky-map update (green). SNR = 6.0, area = $14403~\text{deg}^2$.  Motion begins from $(209.06,6.32)$. & Third sky-map update (green). SNR = 6.0, area = $14403~\text{deg}^2$. Motion begins from $(279.97,-37.45)$. \\
\hline
$-9.5$ s & $41.7$ s & --- & Arrives at first scan position (bottom-left). \\
\hline
$-8.2$ s & $43.0$ s & Arrives at first scan position (top-left). & --- \\
\hline
$-6.5$ s & $44.7$ s & Fourth sky-map update (blue). SNR = 10.5, area = $1029~\text{deg}^2$. Slews toward upper-right of search area. & Fourth sky-map update (blue). SNR = 10.5, area = $1029~\text{deg}^2$. Slews toward bottom-right of search area. \\
\hline
$0.0$ s & $51.2$ s & \multicolumn{2}{|c|}{Merger occurs.} \\
\hline
$+3.2$ s & $54.4$ s & --- & Continues scanning bottom-left region. \\
\hline
$+11.0$ s & $62.2$ s & Begins scanning with three $15~\text{s}$ stops (see Subsec. \ref{Telescope Scanning Model}). & --- \\
\hline
$+51.1$ s & $102.3$ s & Detects event at $(100.54,6.32)$ but does not communicate; halts. & --- \\
\hline
$+100$ s & $151.2$ s & --- & Completes assigned bottom-left search; begins slew toward top-right updated region. \\
\hline
$+136.2$ s & $187.4$ s & --- & Arrives in top-right region; begins scanning. \\
\hline
$+356.8$ s & $408.0$ s & --- & Detects event after extended search, including $15~\text{s}$ integration. \\
\hline
\end{tabular}
\end{table}

\subsection{Detection time histograms}\label{Detection time histograms}
We now compare the distributions of detection times across the three telescope-coordination methods: \textit{Two Step Localization}, \textit{Partial (Limited) Communication}, and \textit{No Communication}. For each simulated event, all source and instrumental parameters are regenerated (see Sec.~\ref{Simulation}), including sky location, orientation, and detector noise realization. The initial pointing of both telescopes is randomized in each iteration, ensuring that variations in detection latency reflect only the coordination strategy and not fixed geometric bias.
In every case, the detection time is measured relative to the merger epoch. The resulting histograms therefore quantify how rapidly each method achieves electromagnetic confirmation of a gravitational-wave event. Black bars denote auxiliary detections in the \textit{Two Step Localization Method} when the auxiliary telescope detects first, subsequently triggering a refined search by the main telescope. The discrete peaks at early times correspond to the four pre-merger sky-map update intervals discussed in Sec.~\ref{Simulation}, reflecting the cadence of localization information available to the telescopes.
Comparing these distributions across LVK runs (O3a, O4, and O5) reveals how improved network sensitivity and geometry influence detection latency. As expected, events in O3a are generally detected more quickly for all methods because the sources were typically closer, leading to higher SNRs and correspondingly smaller localization areas. This trend sets a physical baseline for evaluating the relative efficiency of the different telescope-coordination strategies.

For FOV $=100~\text{deg}^2$, the Two Step Method is consistently the fastest across LVK runs.\\
For LVK O3a, Two Step reaches detection in $95.8$~sec on average (std $135.1$~sec), compared to $196.3$~sec (std $326.2$~sec) for the Limited Communication Method and $247.2$~sec (std $453.6$~sec) for the No Communication Method. This corresponds to roughly a factor $\sim2.0$ improvement over Limited Communication and $\sim2.6$ over No Communication.\\
For LVK O4, Two Step averages $101.0$~sec (std $207.1$~sec), versus $183.6$~sec (std $384.2$~sec) and $243.8$~sec (std $560.0$~sec) for the Limited and No Communication cases, respectively. The speedup is again close to a factor of $2$.\\
For LVK O5, Two Step averages $134.3$~sec (std $158.7$~sec), while Limited and No Communication require $310.9$~sec (std $505.2$~sec) and $395.5$~sec (std $542.2$~sec), respectively. Here the gain is even stronger: faster by a factor of $\sim2.3$ relative to Limited Communication and $\sim2.9$ relative to No Communication.\\
Overall, detection times in O3a are shorter for all methods because event distances in that run were generally smaller, leading to higher SNRs and therefore smaller localization areas.
\begin{figure}[H]
    \centering
    \includegraphics[scale=0.45]{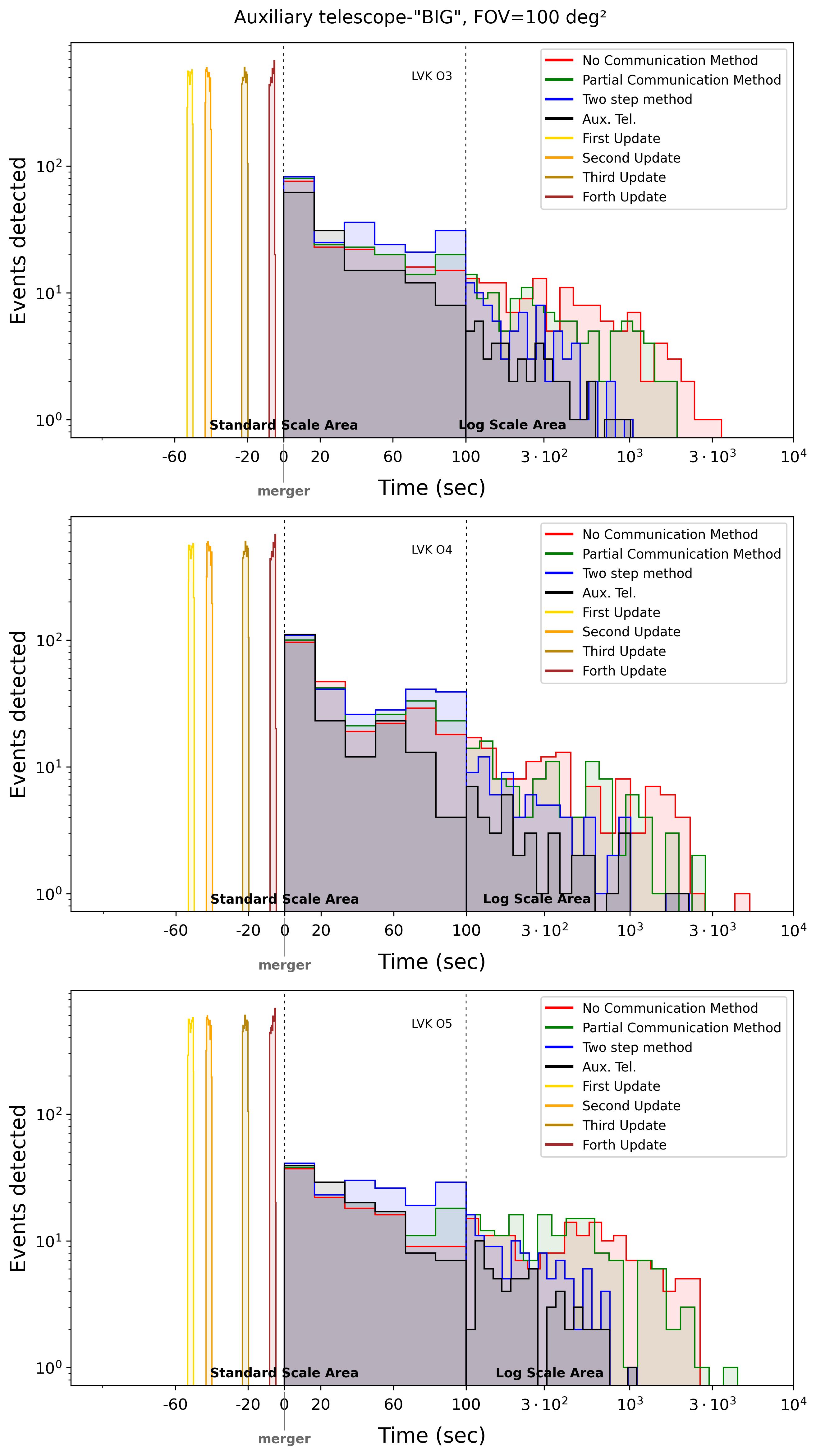}
    \caption{Histogram of detection times with "\textit{BIG}" as the auxiliary telescope with $100 \text{ deg}^2$ FOV.}
\end{figure}
\begin{table}[H]
\centering
\caption{Average detection time and standard deviation for FOV = 100\,deg$^{2}$ across LVK runs.}
\begin{minipage}[t]{0.32\textwidth}
\centering
\textbf{LVK O3a}\\[4pt]
\begin{tabular}{lcc}
\toprule
Method & Avg [sec] & Std [sec] \\
\midrule
Two Step        & 95.8  & 135.1 \\
Limited Com.    & 196.3 & 326.2 \\
No Com.         & 247.2 & 453.6 \\
\bottomrule
\end{tabular}
\end{minipage}
\hfill
\begin{minipage}[t]{0.32\textwidth}
\centering
\textbf{LVK O4}\\[4pt]
\begin{tabular}{lcc}
\toprule
Method & Avg [sec] & Std [sec] \\
\midrule
Two Step        & 101.0 & 207.1 \\
Limited Com.    & 183.6 & 384.2 \\
No Com.         & 243.8 & 560.0 \\
\bottomrule
\end{tabular}
\end{minipage}
\hfill
\begin{minipage}[t]{0.32\textwidth}
\centering
\textbf{LVK O5}\\[4pt]
\begin{tabular}{lcc}
\toprule
Method & Avg [sec] & Std [sec] \\
\midrule
Two Step        & 134.3 & 158.7 \\
Limited Com.    & 310.9 & 505.2 \\
No Com.         & 395.5 & 542.2 \\
\bottomrule
\end{tabular}
\end{minipage}
\end{table}
\begin{figure}[H]
    \centering
    \includegraphics[scale=0.45]{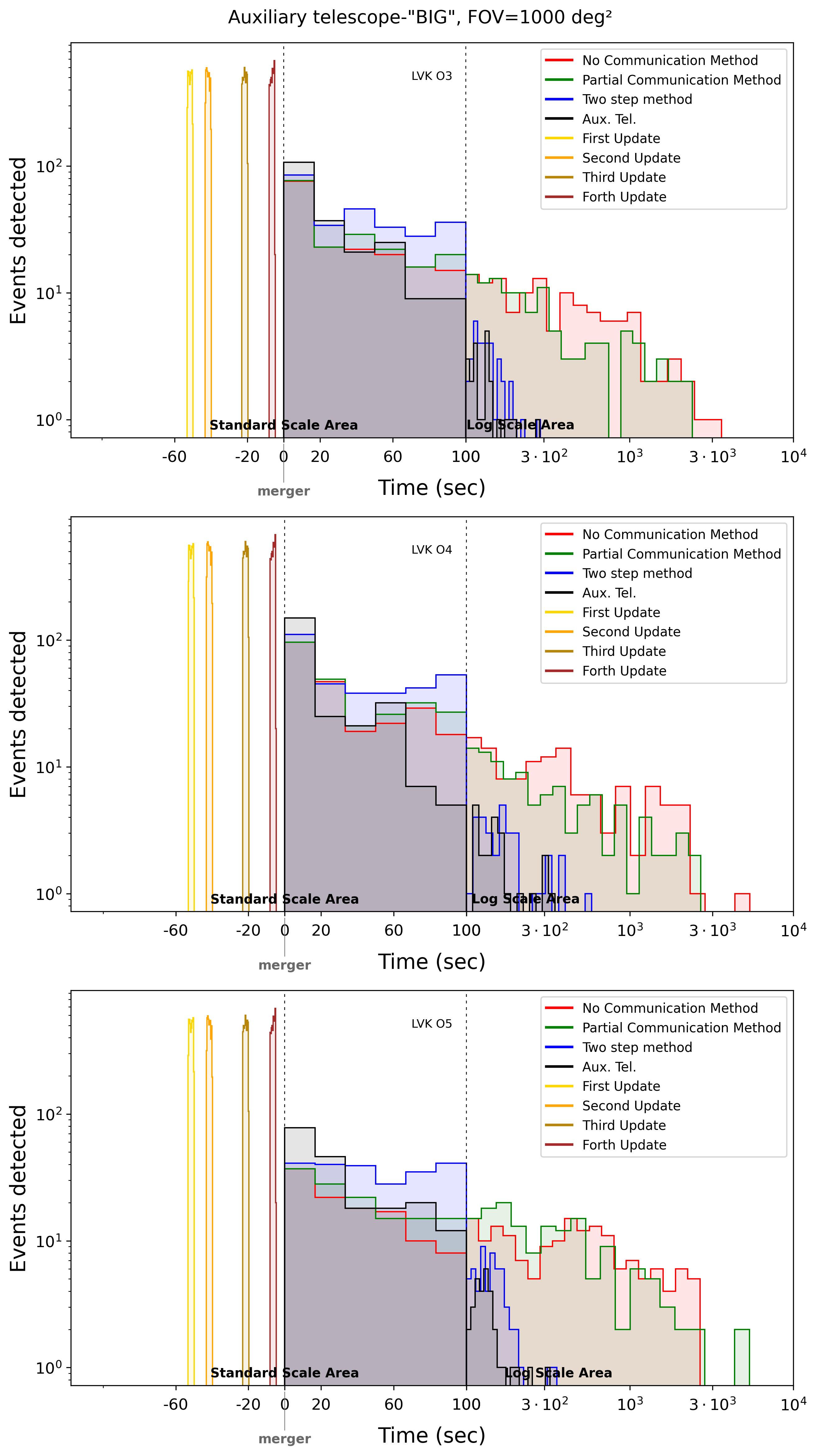}
    \caption{Histogram of detection times with "\textit{BIG}" as the auxiliary telescope with $1000 \text{ deg}^2$ FOV.}
\end{figure}
\begin{table}[H]
\centering
\caption{Average detection time and standard deviation for FOV = 1000\,deg$^{2}$ across LVK runs.}
\begin{minipage}[t]{0.32\textwidth}
\centering
\textbf{LVK O3a}\\[4pt]
\begin{tabular}{lcc}
\toprule
Method & Avg [sec] & Std [sec] \\
\midrule
Two Step        & 53.3  & 47.0 \\
Limited Com.    & 181.7 & 349.4 \\
No Com.         & 237.4 & 439.3 \\
\bottomrule
\end{tabular}
\end{minipage}
\hfill
\begin{minipage}[t]{0.32\textwidth}
\centering
\textbf{LVK O4}\\[4pt]
\begin{tabular}{lcc}
\toprule
Method & Avg [sec] & Std [sec] \\
\midrule
Two Step        & 54.6  & 62.4 \\
Limited Com.    & 167.5 & 359.9 \\
No Com.         & 240.4 & 559.1 \\
\bottomrule
\end{tabular}
\end{minipage}
\hfill
\begin{minipage}[t]{0.32\textwidth}
\centering
\textbf{LVK O5}\\[4pt]
\begin{tabular}{lcc}
\toprule
Method & Avg [sec] & Std [sec] \\
\midrule
Two Step        & 69.2  & 53.0 \\
Limited Com.    & 291.1 & 570.5 \\
No Com.         & 384.7 & 542.5 \\
\bottomrule
\end{tabular}
\end{minipage}
\end{table}

For FOV $=1000~\text{deg}^2$, \textit{Two Step Localization} continues to outperform the other coordination strategies.\\
For LVK~O3a, detection is reached in $53.3$~sec on average (std $47.0$~sec), compared to $181.7$~sec (std $349.4$~sec) for the Limited Communication Method and $237.4$~sec (std $439.3$~sec) for the No Communication Method, giving a factor of $\sim3.4$ and $\sim4.4$ improvement, respectively.\\
For LVK~O4, Two Step averages $54.6$~sec (std $62.4$~sec), while Limited and No Communication reach $167.5$~sec (std $359.9$~sec) and $240.4$~sec (std $559.1$~sec), again showing factors of about $\sim3.1$ and $\sim4.4$ improvement.\\
For LVK~O5, Two Step averages $69.2$~sec (std $53.0$~sec), while Limited and No Communication take $291.1$~sec (std $570.5$~sec) and $384.7$~sec (std $542.5$~sec), yielding factors of $\sim4.2$ and $\sim5.6$ improvement.\\
Overall, O3a events are faster for all methods because the typical source distances were shorter, resulting in higher SNRs and thus smaller localization areas.

\begin{figure}[H]
    \centering
    \includegraphics[scale=0.45]{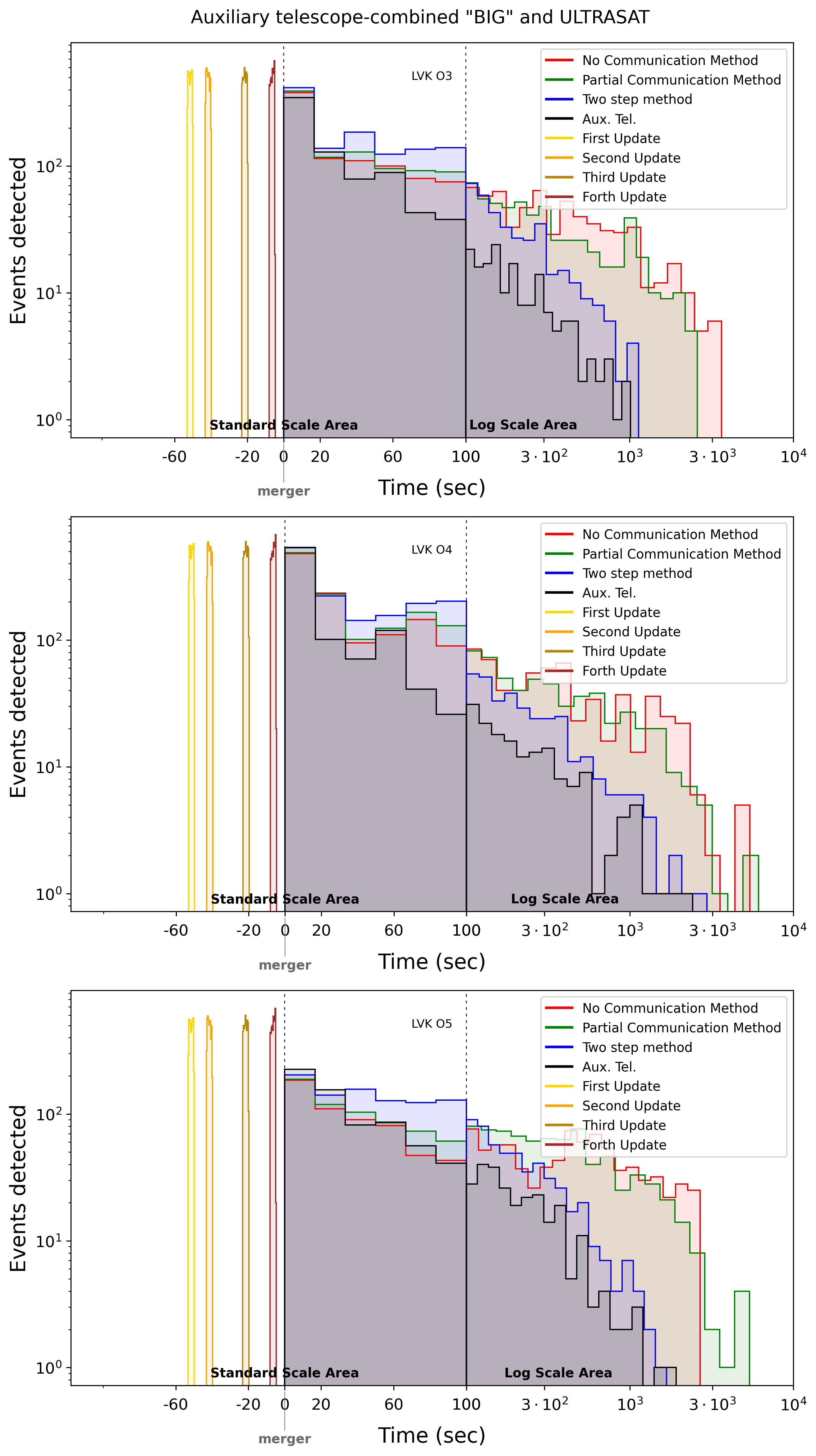}
    \caption{Histogram of detection times with all options of an auxiliary telescope (combined data from all "\textit{BIG}" FOV configurations and from ULTRASAT)}
\end{figure}
\begin{table}[H]
\centering
\caption{Average detection time and standard deviation across all FOVs (100--1000\,deg$^{2}$ and 204\,deg$^{2}$ ULTRASAT equivalent) for each LVK run.}
\begin{minipage}[t]{0.32\textwidth}
\centering
\textbf{LVK O3a}\\[4pt]
\begin{tabular}{lcc}
\toprule
Method & Avg [sec] & Std [sec] \\
\midrule
Two Step        & 87.3~sec  & 115.8~sec \\
Limited Com.    & 195.2~sec & 348.8~sec \\
No Com.         & 244.8~sec & 454.3~sec \\
\bottomrule
\end{tabular}
\end{minipage}
\hfill
\begin{minipage}[t]{0.32\textwidth}
\centering
\textbf{LVK O4}\\[4pt]
\begin{tabular}{lcc}
\toprule
Method & Avg [sec] & Std [sec] \\
\midrule
Two Step        & 89.0~sec  & 157.3~sec \\
Limited Com.    & 178.3~sec & 397.9~sec \\
No Com.         & 244.5~sec & 568.5~sec \\
\bottomrule
\end{tabular}
\end{minipage}
\hfill
\begin{minipage}[t]{0.32\textwidth}
\centering
\textbf{LVK O5}\\[4pt]
\begin{tabular}{lcc}
\toprule
Method & Avg [sec] & Std [sec] \\
\midrule
Two Step        & 123.1~sec & 137.7~sec \\
Limited Com.    & 308.1~sec & 507.2~sec \\
No Com.         & 395.0~sec & 546.0~sec \\
\bottomrule
\end{tabular}
\end{minipage}
\end{table}

Combining all \textit{BIG} FOV configurations and ULTRASAT, \textit{Two Step Localization} remains consistently the fastest across LVK runs.\\
For LVK~O3a, detection occurs in $87.3$~sec on average (std $115.8$~sec), compared with $195.2$~sec (std $348.8$~sec) for the Limited Communication Method and $244.8$~sec (std $454.3$~sec) for the No Communication Method, giving improvements of roughly a factor $\sim2.2$ and $\sim2.8$, respectively.\\
For LVK~O4, Two Step averages $89.0$~sec (std $157.3$~sec), while Limited and No Communication reach $178.3$~sec (std $397.9$~sec) and $244.5$~sec (std $568.5$~sec), maintaining nearly a factor of $2$ advantage.\\
For LVK~O5, Two Step averages $123.1$~sec (std $137.7$~sec), whereas Limited and No Communication require $308.1$~sec (std $507.2$~sec) and $395.0$~sec (std $546.0$~sec), corresponding to factors of $\sim2.5$ and $\sim3.2$ improvement, respectively.\\
Overall, O3a shows shorter detection times for all methods because the events were generally closer, resulting in higher signal-to-noise ratios (SNRs) and hence smaller localization areas. The combined analysis confirms that, across all telescope configurations, \textit{Two Step Localization} consistently achieves substantially faster detections than both \textit{Partial Communication} and \textit{No Communication} methods.

Overall, these results demonstrate that \textit{Two Step Localization} substantially reduces detection time relative to both \textit{Partial Communication Method} and \textit{No Communication Method}, with improvements ranging from about $50\%$ to nearly $80\%$ depending on configuration. Moreover, scaling the auxiliary telescope FOV from $100$ to $1000~\text{deg}^2$ yields nearly a factor-of-two improvement in \textit{Two Step Localization} detection time, showing the strong advantage of wide-field designs in coordinated strategies.

\subsection{NSBH EM–counterpart detection and outlook}\label{NSBH EM Counterpart Detection and Future Directions}
Our simulations do not include NSBH tidal–disruption EM precursors, which are expected to emerge within the final $\sim0.5~\mathrm{ms}$ before coalescence \cite{Tiwari2021}. By construction, our scans begin at the GW–inferred merger time.

Nevertheless, the detection–time distributions (Subsec.~\ref{Detection time histograms}) show that the \textit{Two Step Localization}  method yields a non-negligible fraction of very early confirmations: $1.70\%$ within $0.1~\mathrm{s}$ and $6.16\%$ within $1~\mathrm{s}$ after merger. Hence, a precursor that appears $<1~\mathrm{ms}$ before merger would remain compatible with our operational cadence and follow-up latency.

\section{Summary \& Discussion}\label{sec:summary_discussion}
We investigated how to reduce the latency of EM counterpart detections following BNS mergers identified via GW observations. To this end, we developed a simulation framework that generates localization sky-maps for large ensembles of BNS events and evaluates competing follow-up strategies under controlled conditions.

The key contribution of this work is the development and implementation of the \textit{Two Step Localization} method. This strategy employs an auxiliary telescope and a main telescope. In \textit{Two Step Localization} both the auxiliary and the main telescopes actively survey the evolving GW localization while sharing scan-coverage in real time to avoid duplication. Upon detecting a potential EM counterpart, the auxiliary telescope alerts the main telescope, which then performs a targeted follow-up observation. In \textit{Two Step Localization}, the main telescope can independently detect the event if it is the first to observe it.
We benchmarked \textit{Two Step Localization} against two baselines: the \textit{Partial Communication} strategy, where only scan-coverage is exchanged while detections are not, and the \textit{No Communication} strategy, where no information is shared. Detection-time histograms—constructed by pooling all FOV variants of the \textit{BIG} telescope together with \textit{ULTRASAT} as possible auxiliaries—show that \textit{Two Step Localization} consistently outperforms the baselines. Quantitatively, \textit{Two Step Localization} reduces the mean detection time by $52.61\%$ relative to Partial Communication and by $63.90\%$ relative to No Communication. These results indicate that coordination and timely cross-instrument handoff are effective levers for accelerating EM discovery in multi-messenger campaigns.
Our simulations idealize several aspects of the observational workflow. We model a detection as occurring once the transient exceeds a fixed limiting magnitude in a representative band, without explicitly tracking filter changes, color evolution, weather losses, or detailed scheduling constraints. In reality, wide-field surveys employ real-time pipelines for image subtraction, transient classification, and candidate vetting (e.g.\ ZTF \cite{ZTF}), which introduce additional latencies and modest selection effects. Likewise, the limiting magnitudes we adopt correspond to nominal single-epoch point-source sensitivities and do not fully capture complications such as crowding or host-galaxy surface brightness. The absolute detection times reported here should therefore be interpreted as indicative of the expected timescale rather than precise forecasts for any specific facility. However, these additional effects are expected to impact all follow-up strategies in a broadly similar way, so the relative advantage of \textit{Two-Step Localization} over the baseline strategies should remain robust.

\textit{Scope and outlook.} The present conclusions are based on detection-time statistics from combined simulations across the \textit{BIG} FOV variants and \textit{ULTRASAT} configurations. As detector sensitivity improves and the number of detectors grows, two telescopes may suffice, avoiding delays caused by global communication and coordination lag.

\section{Conclusions \& Future Work}\label{sec:conclusions_future_work}

\textit{Conclusions.} Quantitative outcomes and method comparisons are provided in Sec.~\ref{sec:summary_discussion}. The chief implication is that coordination and streamlined observing operations are pivotal for reducing EM detection latency.
While our simulations employed a hypothetical wide-field facility, denoted \textit{BIG}, which is modeled in a ZTF-like manner with varying FOV settings (100, 200, 400, and 1000~$\mathrm{deg}^2$), a concrete real-world example of such an instrument is the Israeli LAST telescope of the Weizmann Institute at Neot Smadar \cite{Ofek_2023}, which provides a wide-field mode of $355~\mathrm{deg}^2$.
\textit{Future Work.} We outline two focused directions driven by the identified latency sources and by the proximity of detections to coalescence. 

Current observing runs already benefit from networks with three operating GW detectors. Compared to two-detector configurations, three-detector networks yield substantially improved localizations, primarily because arrival-time differences (Shapiro time delays) can be measured more accurately across multiple baselines. GW170817 illustrates both the value and the limitations of such networks: while the presence of Virgo reduced the localization region to $\sim$28~deg$^2$, its sensitivity to the source’s sky position was low, and a transient glitch at LIGO–Livingston complicated the analysis and contributed to the several-hour delay in releasing the first skymap \cite{Abbott2017a}. Even with these challenges, the three-detector configuration enabled a much tighter localization than would have been possible with only Hanford and Livingston. Had all three detectors been fully sensitive to the source, the localization would likely have been even smaller and available at earlier times. This example underscores that robust three-detector observations provide more reliable sky areas and parameter estimates, offering a stronger starting point for methods such as Two Step Localization.

{\bf{EM-Bright Mergers: Priorities for Rapid Detection}\\}

Delays are dominated by field-to-field slews and within-field scan time. A targeted sensitivity study should rank these factors under identical alert streams and sky-map updates.

\noindent\textbf{Open questions.}
\begin{itemize}
    \item What increase in slew rate is required to achieve a specified fractional reduction in time to first detection?
    \item By how much does shortening per-field dwell and adopting faster scan patterns decrease time to first detection, accounting for realistic readout, shutter, and settle times?
    \item Under matched alert/update cadence, which lever most effectively reduces latency: faster slewing, shorter scan time, or improved auxiliary resolution/limiting magnitude?
\end{itemize}

{\bf{NSBH Pre-Merger Detection with the Two Step Localization Method}\\}

Tidal-disruption emission may precede coalescence (see Subsec.~\ref{NSBH EM Counterpart Detection and Future Directions}). Our current scans begin at merger, while detections cluster very near that time (see Subsec.~\ref{Detection time histograms}), suggesting feasibility for capturing signals that turn on shortly before merger with minimal added instrument- and scheduling-related time.

We note that reducing scan times from $\sim 15\,\mathrm{s}$ to the order of $\sim 10\,\mathrm{ms}$ would itself represent a dramatic improvement. In the context of NSBH systems, such a cadence is comparable to the $\sim 0.5\,\mathrm{ms}$ timescale expected for tidal–disruption precursors \cite{Tiwari2021}, and would therefore make the capture of these extremely short-lived signals feasible. This leap would fundamentally alter the achievable cadence, bringing sub-second pre-merger detections well within reach, even without major changes to slewing or sky-map update rates.

\noindent\textbf{Open questions.}
\begin{itemize}
    \item If the auxiliary records frames during the final instants before the predicted merger, how does the pre-merger capture rate change?
    \item For NSBH events, what reduction in detection time is expected from faster slewing relative to the present setup?
    \item For NSBH events, does higher auxiliary resolution with better limiting magnitude reduce latency more effectively than faster slewing?
\end{itemize}

\section*{Acknowledgments}\label{section: Acknowledgments}
We acknowledge support from the US-Israel Binational Science Fund (BSF) grant No. 2020245 and the Israel Science Fund (ISF) grant No. 1698/22. \\
In addition, This research has made use of data or software obtained from the Gravitational Wave Open Science Center (gwosc.org), a service of the LIGO Scientific Collaboration, the Virgo Collaboration, and KAGRA. This material is based upon work supported by NSF's LIGO Laboratory which is a major facility fully funded by the National Science Foundation, as well as the Science and Technology Facilities Council (STFC) of the United Kingdom, the Max-Planck-Society (MPS), and the State of Niedersachsen/Germany for support of the construction of Advanced LIGO and construction and operation of the GEO600 detector. Additional support for Advanced LIGO was provided by the Australian Research Council. Virgo is funded, through the European Gravitational Observatory (EGO), by the French Centre National de Recherche Scientifique (CNRS), the Italian Istituto Nazionale di Fisica Nucleare (INFN) and the Dutch Nikhef, with contributions by institutions from Belgium, Germany, Greece, Hungary, Ireland, Japan, Monaco, Poland, Portugal, Spain. KAGRA is supported by Ministry of Education, Culture, Sports, Science and Technology (MEXT), Japan Society for the Promotion of Science (JSPS) in Japan; National Research Foundation (NRF) and Ministry of Science and ICT (MSIT) in Korea; Academia Sinica (AS) and National Science and Technology Council (NSTC) in Taiwan.


\begin{thebibliography}{34}%
\makeatletter
\providecommand \@ifxundefined [1]{%
 \@ifx{#1\undefined}
}%
\providecommand \@ifnum [1]{%
 \ifnum #1\expandafter \@firstoftwo
 \else \expandafter \@secondoftwo
 \fi
}%
\providecommand \@ifx [1]{%
 \ifx #1\expandafter \@firstoftwo
 \else \expandafter \@secondoftwo
 \fi
}%
\providecommand \natexlab [1]{#1}%
\providecommand \enquote  [1]{``#1''}%
\providecommand \bibnamefont  [1]{#1}%
\providecommand \bibfnamefont [1]{#1}%
\providecommand \citenamefont [1]{#1}%
\providecommand \href@noop [0]{\@secondoftwo}%
\providecommand \href [0]{\begingroup \@sanitize@url \@href}%
\providecommand \@href[1]{\@@startlink{#1}\@@href}%
\providecommand \@@href[1]{\endgroup#1\@@endlink}%
\providecommand \@sanitize@url [0]{\catcode `\\12\catcode `\$12\catcode `\&12\catcode `\#12\catcode `\^12\catcode `\_12\catcode `\%12\relax}%
\providecommand \@@startlink[1]{}%
\providecommand \@@endlink[0]{}%
\providecommand \url  [0]{\begingroup\@sanitize@url \@url }%
\providecommand \@url [1]{\endgroup\@href {#1}{\urlprefix }}%
\providecommand \urlprefix  [0]{URL }%
\providecommand \Eprint [0]{\href }%
\providecommand \doibase [0]{https://doi.org/}%
\providecommand \selectlanguage [0]{\@gobble}%
\providecommand \bibinfo  [0]{\@secondoftwo}%
\providecommand \bibfield  [0]{\@secondoftwo}%
\providecommand \translation [1]{[#1]}%
\providecommand \BibitemOpen [0]{}%
\providecommand \bibitemStop [0]{}%
\providecommand \bibitemNoStop [0]{.\EOS\space}%
\providecommand \EOS [0]{\spacefactor3000\relax}%
\providecommand \BibitemShut  [1]{\csname bibitem#1\endcsname}%
\let\auto@bib@innerbib\@empty
 
\bibitem [{\citenamefont {Abbott}\ \emph {et~al.}(2017{\natexlab{a}})\citenamefont {Abbott} \emph {et~al.}}]{Abbott2017a}%
  \BibitemOpen
  \bibfield  {author} {\bibinfo {author} {\bibfnamefont {B.~P.}\ \bibnamefont {Abbott}} \emph {et~al.} (\bibinfo {collaboration} {LIGO Scientific, Virgo}),\ }\bibfield  {title} {\bibinfo {title} {{GW170817: Observation of Gravitational Waves from a Binary Neutron Star Inspiral}},\ }\href {https://doi.org/10.1103/PhysRevLett.119.161101} {\bibfield  {journal} {\bibinfo  {journal} {Phys. Rev. Lett.}\ }\textbf {\bibinfo {volume} {119}},\ \bibinfo {pages} {161101} (\bibinfo {year} {2017}{\natexlab{a}})},\ \Eprint {https://arxiv.org/abs/1710.05832} {arXiv:1710.05832 [gr-qc]} \BibitemShut {NoStop}%
  
\bibitem [{\citenamefont {Abbott}\ \emph {et~al.}(2017{\natexlab{c}})\citenamefont {Abbott} \emph {et~al.}}]{Kilonova2017}%
  \BibitemOpen
  \bibfield  {author} {\bibinfo {author} {\bibfnamefont {B.~P.}\ \bibnamefont {Abbott}} \emph {et~al.} (\bibinfo {collaboration} {LIGO Scientific, Virgo}),\ }\bibfield  {title} {\bibinfo {title} {{Estimating the Contribution of Dynamical Ejecta in the Kilonova Associated with GW170817}},\ }\href {https://doi.org/10.3847/2041-8213/aa9478} {\bibfield  {journal} {\bibinfo  {journal} {Astrophys. J. Lett.}\ }\textbf {\bibinfo {volume} {850}},\ \bibinfo {pages} {L39} (\bibinfo {year} {2017}{\natexlab{c}})},\ \Eprint {https://arxiv.org/abs/1710.05836} {arXiv:1710.05836 [astro-ph.HE]} \BibitemShut {NoStop}%

  
\bibitem [{\citenamefont {Abbott}\ \emph {et~al.}(2017{\natexlab{d}})\citenamefont {Abbott} \emph {et~al.}}]{H02017}%
  \BibitemOpen
  \bibfield  {author} {\bibinfo {author} {\bibfnamefont {B.~P.}\ \bibnamefont {Abbott}} \emph {et~al.} (\bibinfo {collaboration} {LIGO Scientific, Virgo, 1M2H, Dark Energy Camera GW-E, DES, DLT40, Las Cumbres Observatory, VINROUGE, MASTER}),\ }\bibfield  {title} {\bibinfo {title} {{A gravitational-wave standard siren measurement of the Hubble constant}},\ }\href {https://doi.org/10.1038/nature24471} {\bibfield  {journal} {\bibinfo  {journal} {Nature}\ }\textbf {\bibinfo {volume} {551}},\ \bibinfo {pages} {85} (\bibinfo {year} {2017}{\natexlab{d}})},\ \Eprint {https://arxiv.org/abs/1710.05835} {arXiv:1710.05835 [astro-ph.CO]} \BibitemShut {NoStop}%
 \bibitem [{\citenamefont {Abbott}\ \emph {et~al.}(2017{\natexlab{e}})\citenamefont {Abbott} \emph {et~al.}}]{GRB2017}%
  \BibitemOpen
  \bibfield  {author} {\bibinfo {author} {\bibfnamefont {B.~P.}\ \bibnamefont {Abbott}} \emph {et~al.} (\bibinfo {collaboration} {LIGO Scientific, Virgo, Fermi-GBM, INTEGRAL}),\ }\bibfield  {title} {\bibinfo {title} {{Gravitational Waves and Gamma-rays from a Binary Neutron Star Merger: GW170817 and GRB 170817A}},\ }\href {https://doi.org/10.3847/2041-8213/aa920c} {\bibfield  {journal} {\bibinfo  {journal} {Astrophys. J. Lett.}\ }\textbf {\bibinfo {volume} {848}},\ \bibinfo {pages} {L13} (\bibinfo {year} {2017}{\natexlab{e}})},\ \Eprint {https://arxiv.org/abs/1710.05834} {arXiv:1710.05834 [astro-ph.HE]} \BibitemShut {NoStop}%
  \bibitem [{\citenamefont {Coughlin}\ \emph {et~al.}(2019)\citenamefont
  {Coughlin} \emph {et~al.}}]{Coughlin:2018fis}%
  \BibitemOpen
  \bibfield  {author} {\bibinfo {author} {\bibfnamefont {M.~W.}\ \bibnamefont
  {Coughlin}} \emph {et~al.},\ }\bibfield  {title} {\bibinfo {title}
  {{Multimessenger Bayesian parameter inference of a binary neutron star merger}},\ }\href
  {https://doi.org/10.1093/mnrasl/slz133} {\bibfield  {journal} {\bibinfo
  {journal} {Mon. Not. R. Astron. Soc.}\ }\textbf {\bibinfo {volume} {489}},\
  \bibinfo {number} {1}\ \bibinfo {pages} {L91--L96} (\bibinfo {year} {2019})},\
  \Eprint {https://arxiv.org/abs/1812.04803} {arXiv:1812.04803 [astro-ph.HE]}%
  \BibitemShut {NoStop}%

  \bibitem [{\citenamefont {Margutti}\ \emph {et~al.}(2017)\citenamefont
  {Margutti} \emph {et~al.}}]{Margutti:2017cjl}%
  \BibitemOpen
  \bibfield  {author} {\bibinfo {author} {\bibfnamefont {R.}\ \bibnamefont
  {Margutti}} \emph {et~al.},\ }\bibfield  {title} {\bibinfo {title} {{The Electromagnetic Counterpart of the Binary Neutron Star Merger LIGO/VIRGO GW170817. V. Rising X-ray Emission from an Off-Axis Jet}},\ }\href
  {https://doi.org/10.3847/2041-8213/aa9057} {\bibfield  {journal} {\bibinfo
  {journal} {Astrophys. J. Lett.}\ }\textbf {\bibinfo {volume} {848}},\ \bibinfo
  {number} {2}\ \bibinfo {pages} {L20} (\bibinfo {year} {2017})},\ \Eprint
  {https://arxiv.org/abs/1710.05431} {arXiv:1710.05431 [astro-ph.HE]}%
  \BibitemShut {NoStop}%

\bibitem [{\citenamefont {Shappee}\ \emph {et~al.}(2017)\citenamefont
  {Shappee} \emph {et~al.}}]{Shappee:2017zly}%
  \BibitemOpen
  \bibfield  {author} {\bibinfo {author} {\bibfnamefont {B.~J.}\ \bibnamefont
  {Shappee}} \emph {et~al.},\ }\bibfield  {title} {\bibinfo {title} {{Early Spectra of the Gravitational Wave Source GW170817: Evolution of a Neutron Star Merger}},\ }\href
  {https://doi.org/10.1126/science.aaq0186} {\bibfield  {journal} {\bibinfo
  {journal} {Science}\ }\textbf {\bibinfo {volume} {358}},\ \bibinfo {pages}
  {1574} (\bibinfo {year} {2017})},\ \Eprint {https://arxiv.org/abs/1710.05432}
  {arXiv:1710.05432 [astro-ph.HE]}%
  \BibitemShut {NoStop}%

\bibitem [{\citenamefont {Troja}\ \emph {et~al.}(2017)\citenamefont
  {Troja} \emph {et~al.}}]{Troja:2017nqp}%
  \BibitemOpen
  \bibfield  {author} {\bibinfo {author} {\bibfnamefont {E.}\ \bibnamefont
  {Troja}} \emph {et~al.},\ }\bibfield  {title} {\bibinfo {title} {{The X-ray counterpart to the gravitational wave event GW 170817}},\ }\href
  {https://doi.org/10.1038/nature24290} {\bibfield  {journal} {\bibinfo
  {journal} {Nature}\ }\textbf {\bibinfo {volume} {551}},\ \bibinfo {pages}
  {71--74} (\bibinfo {year} {2017})},\ \Eprint {https://arxiv.org/abs/1710.05433}
  {arXiv:1710.05433 [astro-ph.HE]}%
  \BibitemShut {NoStop}%

\bibitem [{\citenamefont {Kilpatrick}\ \emph {et~al.}(2017)\citenamefont
  {Kilpatrick} \emph {et~al.}}]{Kilpatrick:2017mhz}%
  \BibitemOpen
  \bibfield  {author} {\bibinfo {author} {\bibfnamefont {C.~D.}\ \bibnamefont
  {Kilpatrick}} \emph {et~al.},\ }\bibfield  {title} {\bibinfo {title} {{Electromagnetic Evidence that SSS17a is the Result of a Binary Neutron Star Merger}},\ }\href
  {https://doi.org/10.1126/science.aaq0073} {\bibfield  {journal} {\bibinfo
  {journal} {Science}\ }\textbf {\bibinfo {volume} {358}},\ \bibinfo {number}
  {6370}\ \bibinfo {pages} {1583--1587} (\bibinfo {year} {2017})},\ \Eprint
  {https://arxiv.org/abs/1710.05434} {arXiv:1710.05434 [astro-ph.HE]}%
  \BibitemShut {NoStop}%

\bibitem [{\citenamefont {Hallinan}\ \emph {et~al.}(2017)\citenamefont
  {Hallinan} \emph {et~al.}}]{Hallinan:2017woc}%
  \BibitemOpen
  \bibfield  {author} {\bibinfo {author} {\bibfnamefont {G.}\ \bibnamefont
  {Hallinan}} \emph {et~al.},\ }\bibfield  {title} {\bibinfo {title} {{A Radio Counterpart to a Neutron Star Merger}},\ }\href
  {https://doi.org/10.1126/science.aap9855} {\bibfield  {journal} {\bibinfo
  {journal} {Science}\ }\textbf {\bibinfo {volume} {358}},\ \bibinfo {pages}
  {1579} (\bibinfo {year} {2017})},\ \Eprint {https://arxiv.org/abs/1710.05435}
  {arXiv:1710.05435 [astro-ph.HE]}%
  \BibitemShut {NoStop}%

\bibitem [{\citenamefont {Kasliwal}\ \emph {et~al.}(2017)\citenamefont
  {Kasliwal} \emph {et~al.}}]{Kasliwal:2017ngb}%
  \BibitemOpen
  \bibfield  {author} {\bibinfo {author} {\bibfnamefont {M.~M.}\ \bibnamefont
  {Kasliwal}} \emph {et~al.},\ }\bibfield  {title} {\bibinfo {title} {{Illuminating Gravitational Waves: A Concordant Picture of Photons from a Neutron Star Merger}},\ }\href
  {https://doi.org/10.1126/science.aap9455} {\bibfield  {journal} {\bibinfo
  {journal} {Science}\ }\textbf {\bibinfo {volume} {358}},\ \bibinfo {pages}
  {1559} (\bibinfo {year} {2017})},\ \Eprint {https://arxiv.org/abs/1710.05436}
  {arXiv:1710.05436 [astro-ph.HE]}%
  \BibitemShut {NoStop}%

\bibitem [{\citenamefont {Evans}\ \emph {et~al.}(2017)\citenamefont
  {Evans} \emph {et~al.}}]{Evans:2017mmy}%
  \BibitemOpen
  \bibfield  {author} {\bibinfo {author} {\bibfnamefont {P.~A.}\ \bibnamefont
  {Evans}} \emph {et~al.},\ }\bibfield  {title} {\bibinfo {title} {{Swift and NuSTAR observations of GW170817: detection of a blue kilonova}},\ }\href
  {https://doi.org/10.1126/science.aap9580} {\bibfield  {journal} {\bibinfo
  {journal} {Science}\ }\textbf {\bibinfo {volume} {358}},\ \bibinfo {pages}
  {1565} (\bibinfo {year} {2017})},\ \Eprint {https://arxiv.org/abs/1710.05437}
  {arXiv:1710.05437 [astro-ph.HE]}%
  \BibitemShut {NoStop}%

\bibitem [{\citenamefont {Drout}\ \emph {et~al.}(2017)\citenamefont
  {Drout} \emph {et~al.}}]{Drout:2017ijr}%
  \BibitemOpen
  \bibfield  {author} {\bibinfo {author} {\bibfnamefont {M.~R.}\ \bibnamefont
  {Drout}} \emph {et~al.},\ }\bibfield  {title} {\bibinfo {title} {{Light Curves of the Neutron Star Merger GW170817/SSS17a: Implications for R-Process Nucleosynthesis}},\ }\href
  {https://doi.org/10.1126/science.aaq0049} {\bibfield  {journal} {\bibinfo
  {journal} {Science}\ }\textbf {\bibinfo {volume} {358}},\ \bibinfo {pages}
  {1570--1574} (\bibinfo {year} {2017})},\ \Eprint
  {https://arxiv.org/abs/1710.05443} {arXiv:1710.05443 [astro-ph.HE]}%
  \BibitemShut {NoStop}%

\bibitem [{\citenamefont {Goldstein}\ \emph {et~al.}(2017)\citenamefont
  {Goldstein} \emph {et~al.}}]{Goldstein:2017mmi}%
  \BibitemOpen
  \bibfield  {author} {\bibinfo {author} {\bibfnamefont {A.}\ \bibnamefont
  {Goldstein}} \emph {et~al.},\ }\bibfield  {title} {\bibinfo {title} {{An Ordinary Short Gamma-Ray Burst with Extraordinary Implications: Fermi-GBM Detection of GRB 170817A}},\ }\href
  {https://doi.org/10.3847/2041-8213/aa8f41} {\bibfield  {journal} {\bibinfo
  {journal} {Astrophys. J. Lett.}\ }\textbf {\bibinfo {volume} {848}},\ \bibinfo
  {number} {2}\ \bibinfo {pages} {L14} (\bibinfo {year} {2017})},\ \Eprint
  {https://arxiv.org/abs/1710.05446} {arXiv:1710.05446 [astro-ph.HE]}%
  \BibitemShut {NoStop}%

\bibitem [{\citenamefont {Pozanenko}\ \emph {et~al.}(2018)\citenamefont
  {Pozanenko} \emph {et~al.}}]{Pozanenko:2017jrn}%
  \BibitemOpen
  \bibfield  {author} {\bibinfo {author} {\bibfnamefont {A.}\ \bibnamefont
  {Pozanenko}} \emph {et~al.},\ }\bibfield  {title} {\bibinfo {title} {{GRB 170817A Associated with GW170817: Multi-frequency Observations and Modeling of Prompt Gamma-Ray Emission}},\ }\href
  {https://doi.org/10.3847/2041-8213/aaa2f6} {\bibfield  {journal} {\bibinfo
  {journal} {Astrophys. J. Lett.}\ }\textbf {\bibinfo {volume} {852}},\ \bibinfo
  {number} {2}\ \bibinfo {pages} {L30} (\bibinfo {year} {2018})},\ \Eprint
  {https://arxiv.org/abs/1710.05448} {arXiv:1710.05448 [astro-ph.HE]}%
  \BibitemShut {NoStop}%

\bibitem [{\citenamefont {Savchenko}\ \emph {et~al.}(2017)\citenamefont
  {Savchenko} \emph {et~al.}}]{Savchenko:2017ffs}%
  \BibitemOpen
  \bibfield  {author} {\bibinfo {author} {\bibfnamefont {V.}\ \bibnamefont
  {Savchenko}} \emph {et~al.},\ }\bibfield  {title} {\bibinfo {title} {{INTEGRAL Detection of the First Prompt Gamma-Ray Signal Coincident with the Gravitational-wave Event GW170817}},\ }\href
  {https://doi.org/10.3847/2041-8213/aa8f94} {\bibfield  {journal} {\bibinfo
  {journal} {Astrophys. J. Lett.}\ }\textbf {\bibinfo {volume} {848}},\ \bibinfo
  {number} {2}\ \bibinfo {pages} {L15} (\bibinfo {year} {2017})},\ \Eprint
  {https://arxiv.org/abs/1710.05449} {arXiv:1710.05449 [astro-ph.HE]}%
  \BibitemShut {NoStop}%

\bibitem [{\citenamefont {Kocevski}\ \emph {et~al.}(2017)\citenamefont
  {Kocevski} \emph {et~al.}}]{Kocevski:2017liw}%
  \BibitemOpen
  \bibfield  {author} {\bibinfo {author} {\bibfnamefont {D.}\ \bibnamefont
  {Kocevski}} \emph {et~al.} (Fermi-LAT),\ }\bibfield  {title} {\bibinfo {title} {{Fermi-LAT observations of the LIGO/Virgo event GW170817}},\ }\bibfield  {journal} {\bibinfo  {journal} {arXiv preprint}\ }\textbf {\bibinfo {volume} {1710.05450}} (\bibinfo {year} {2017}),\ \Eprint
  {https://arxiv.org/abs/1710.05450} {arXiv:1710.05450 [astro-ph.HE]}%
  \BibitemShut {NoStop}%

\bibitem [{\citenamefont {Chornock}\ \emph {et~al.}(2017)\citenamefont
  {Chornock} \emph {et~al.}}]{Chornock:2017sdf}%
  \BibitemOpen
  \bibfield  {author} {\bibinfo {author} {\bibfnamefont {R.}\ \bibnamefont
  {Chornock}} \emph {et~al.},\ }\bibfield  {title}
  {\bibinfo {title} {{The Electromagnetic Counterpart of the Binary Neutron Star Merger LIGO/VIRGO GW170817. IV. Detection of Near-infrared Signatures of r-process Nucleosynthesis with Gemini-South}},\ }\href
  {https://doi.org/10.3847/2041-8213/aa905c} {\bibfield  {journal} {\bibinfo
  {journal} {Astrophys. J. Lett.}\ }\textbf {\bibinfo {volume} {848}},\
  \bibinfo {number} {2}\ \bibinfo {pages} {L19} (\bibinfo {year} {2017})},\
  \Eprint {https://arxiv.org/abs/1710.05454} {arXiv:1710.05454 [astro-ph.HE]}%
  \BibitemShut {NoStop}%

\bibitem [{\citenamefont {Nicholl}\ \emph {et~al.}(2017)\citenamefont
  {Nicholl} \emph {et~al.}}]{Nicholl:2017ahq}%
  \BibitemOpen
  \bibfield  {author} {\bibinfo {author} {\bibfnamefont {M.}\ \bibnamefont
  {Nicholl}} \emph {et~al.},\ }\bibfield  {title}
  {\bibinfo {title} {{The Electromagnetic Counterpart of the Binary Neutron Star Merger LIGO/VIRGO GW170817. III. Optical and UV Spectra of a Blue Kilonova From Fast Polar Ejecta}},\ }\href
  {https://doi.org/10.3847/2041-8213/aa9029} {\bibfield  {journal} {\bibinfo
  {journal} {Astrophys. J. Lett.}\ }\textbf {\bibinfo {volume} {848}},\ 
  \bibinfo {number} {2}\ \bibinfo {pages} {L18} (\bibinfo {year} {2017})},\ 
  \Eprint {https://arxiv.org/abs/1710.05456} {arXiv:1710.05456 [astro-ph.HE]}%
  \BibitemShut {NoStop}%

\bibitem [{\citenamefont {Alexander}\ \emph {et~al.}(2017)\citenamefont
  {Alexander} \emph {et~al.}}]{Alexander:2017aly}%
  \BibitemOpen
  \bibfield  {author} {\bibinfo {author} {\bibfnamefont {K.~D.}\ \bibnamefont
  {Alexander}} \emph {et~al.},\ }\bibfield  {title} {\bibinfo {title} {{The Electromagnetic Counterpart of the Binary Neutron Star Merger LIGO/VIRGO GW170817. VI. Radio Constraints on a Relativistic Jet and Predictions for Late-Time Emission from the Kilonova Ejecta}},\ }\href
  {https://doi.org/10.3847/2041-8213/aa905d} {\bibfield  {journal} {\bibinfo
  {journal} {Astrophys. J. Lett.}\ }\textbf {\bibinfo {volume} {848}},\ \bibinfo
  {number} {2}\ \bibinfo {pages} {L21} (\bibinfo {year} {2017})},\ \Eprint
  {https://arxiv.org/abs/1710.05457} {arXiv:1710.05457 [astro-ph.HE]}%
  \BibitemShut {NoStop}%

\bibitem [{\citenamefont {Soares-Santos}\ \emph {et~al.}(2017)\citenamefont
  {Soares-Santos} \emph {et~al.}}]{DES:2017kbs}%
  \BibitemOpen
  \bibfield  {author} {\bibinfo {author} {\bibfnamefont {M.}~\bibnamefont
  {Soares-Santos}} \emph {et~al.} (DES, Dark Energy Camera GW-EM),\ }\bibfield  {title}
  {\bibinfo {title} {{The Electromagnetic Counterpart of the Binary Neutron Star Merger LIGO/Virgo GW170817. I. Discovery of the Optical Counterpart Using the Dark Energy Camera}},\ }\href
  {https://doi.org/10.3847/2041-8213/aa9059} {\bibfield  {journal} {\bibinfo
  {journal} {Astrophys. J. Lett.}\ }\textbf {\bibinfo {volume} {848}},\ \bibinfo
  {number} {2}\ \bibinfo {pages} {L16} (\bibinfo {year} {2017})},\ \Eprint
  {https://arxiv.org/abs/1710.05459} {arXiv:1710.05459 [astro-ph.HE]}%
  \BibitemShut {NoStop}%

\bibitem [{\citenamefont {Abbott}\ \emph {et~al.}(2017)\citenamefont
  {Abbott} \emph {et~al.}}]{LIGOScientific:2017ync}%
  \BibitemOpen
  \bibfield  {author} {\bibinfo {author} {\bibfnamefont {B.~P.}\ \bibnamefont
  {Abbott}} \emph {et~al.} (LIGO Scientific, Virgo, Fermi GBM, INTEGRAL, IceCube, AstroSat Cadmium Zinc Telluride Imager Team, IPN, Insight-Hxmt, ANTARES, Swift, AGILE Team, 1M2H Team, Dark Energy Camera GW-EM, DES, DLT40, GRAWITA, Fermi-LAT, ATCA, ASKAP, Las Cumbres Observatory Group, OzGrav, DWF, AST3, CAASTRO, VINROUGE, MASTER, J-GEM, GROWTH, JAGWAR, Caltech-NRAO, TTU-NRAO, NuSTAR, Pan-STARRS, MAXI Team, TZAC Consortium, KU, Nordic Optical Telescope, ePESSTO, GROND, Texas Tech University, SALT Group, TOROS, BOOTES, MWA, CALET, IKI-GW Follow-up, H.E.S.S., LOFAR, LWA, HAWC, Pierre Auger, ALMA, Euro VLBI Team, Pi of the Sky, Chandra Team at McGill University, DFN, ATLAS Telescopes, High Time Resolution Universe Survey, RIMAS, RATIR, SKA South Africa/MeerKAT),\ }\bibfield  {title}
  {\bibinfo {title} {{Multi-messenger Observations of a Binary Neutron Star Merger}},\ }\href
  {https://doi.org/10.3847/2041-8213/aa91c9} {\bibfield  {journal} {\bibinfo
  {journal} {Astrophys. J. Lett.}\ }\textbf {\bibinfo {volume} {848}},\ \bibinfo
  {number} {2}\ \bibinfo {pages} {L12} (\bibinfo {year} {2017})},\ \Eprint
  {https://arxiv.org/abs/1710.05833} {arXiv:1710.05833 [astro-ph.HE]}%
  \BibitemShut {NoStop}%

\bibitem [{\citenamefont {Cowperthwaite}\ \emph {et~al.}(2017)\citenamefont
  {Cowperthwaite} \emph {et~al.}}]{Cowperthwaite:2017dyu}%
  \BibitemOpen
  \bibfield  {author} {\bibinfo {author} {\bibfnamefont {P.~S.}\ \bibnamefont
  {Cowperthwaite}} \emph {et~al.},\ }\bibfield  {title}
  {\bibinfo {title} {{The Electromagnetic Counterpart of the Binary Neutron Star Merger LIGO/Virgo GW170817. II. UV, Optical, and Near-infrared Light Curves and Comparison to Kilonova Models}},\ }\href
  {https://doi.org/10.3847/2041-8213/aa8fc7} {\bibfield  {journal} {\bibinfo
  {journal} {Astrophys. J. Lett.}\ }\textbf {\bibinfo {volume} {848}},\ 
  \bibinfo {number} {2}\ \bibinfo {pages} {L17} (\bibinfo {year} {2017})},\ 
  \Eprint {https://arxiv.org/abs/1710.05840} {arXiv:1710.05840 [astro-ph.HE]}%
  \BibitemShut {NoStop}%

\bibitem [{\citenamefont {Siegel}\ and\ \citenamefont {Ciolfi}(2016)}]{Siegel_2016}%
  \BibitemOpen
  \bibfield  {author} {\bibinfo {author} {\bibfnamefont {D.~M.}\ \bibnamefont {Siegel}}\ and\ \bibinfo {author} {\bibfnamefont {R.}~\bibnamefont {Ciolfi}},\ }\bibfield  {title} {\bibinfo {title} {{Electromagnetic emission from long-lived binary neutron star merger remnants II: lightcurves and spectra}},\ }\href {https://doi.org/10.3847/0004-637X/819/1/15} {\bibfield  {journal} {\bibinfo  {journal} {Astrophys. J.}\ }\textbf {\bibinfo {volume} {819}},\ \bibinfo {pages} {15} (\bibinfo {year} {2016})},\ \Eprint {https://arxiv.org/abs/1508.07939} {arXiv:1508.07939 [astro-ph.HE]} \BibitemShut {NoStop}%
\bibitem [{\citenamefont {Fong}\ \emph {et~al.}(2017)\citenamefont {Fong} \emph {et~al.}}]{2017ApJ...848L..23F}%
  \BibitemOpen
  \bibfield  {author} {\bibinfo {author} {\bibfnamefont {W.}~\bibnamefont {Fong}} \emph {et~al.},\ }\bibfield  {title} {\bibinfo {title} {{The Electromagnetic Counterpart of the Binary Neutron Star Merger LIGO/VIRGO GW170817. VIII. A Comparison to Cosmological Short-duration Gamma-ray Bursts}},\ }\href {https://doi.org/10.3847/2041-8213/aa9018} {\bibfield  {journal} {\bibinfo  {journal} {Astrophys. J. Lett.}\ }\textbf {\bibinfo {volume} {848}},\ \bibinfo {pages} {L23} (\bibinfo {year} {2017})},\ \Eprint {https://arxiv.org/abs/1710.05438} {arXiv:1710.05438 [astro-ph.HE]} \BibitemShut {NoStop}%
\bibitem [{\citenamefont {Most}\ and\ \citenamefont {Philippov}(2020)}]{2020ApJ...893L...6M}%
  \BibitemOpen
  \bibfield  {author} {\bibinfo {author} {\bibfnamefont {E.~R.}\ \bibnamefont {Most}}\ and\ \bibinfo {author} {\bibfnamefont {A.~A.}\ \bibnamefont {Philippov}},\ }\bibfield  {title} {\bibinfo {title} {{Electromagnetic precursors to gravitational wave events: Numerical simulations of flaring in pre-merger binary neutron star magnetospheres}},\ }\href {https://doi.org/10.3847/2041-8213/ab8196} {\bibfield  {journal} {\bibinfo  {journal} {Astrophys. J. Lett.}\ }\textbf {\bibinfo {volume} {893}},\ \bibinfo {pages} {L6} (\bibinfo {year} {2020})},\ \Eprint {https://arxiv.org/abs/2001.06037} {arXiv:2001.06037 [astro-ph.HE]} \BibitemShut {NoStop}%
\bibitem [{\citenamefont {Tiwari}\ \emph {et~al.}(2021)\citenamefont {Tiwari}, \citenamefont {Ebersold},\ and\ \citenamefont {Hamilton}}]{Tiwari2021}%
  \BibitemOpen
  \bibfield  {author} {\bibinfo {author} {\bibfnamefont {S.}~\bibnamefont {Tiwari}}, \bibinfo {author} {\bibfnamefont {M.}~\bibnamefont {Ebersold}},\ and\ \bibinfo {author} {\bibfnamefont {E.~Z.}\ \bibnamefont {Hamilton}},\ }\bibfield  {title} {\bibinfo {title} {{Leveraging gravitational-wave memory to distinguish neutron star-black hole binaries from black hole binaries}},\ }\href {https://doi.org/10.1103/PhysRevD.104.123024} {\bibfield  {journal} {\bibinfo  {journal} {Phys. Rev. D}\ }\textbf {\bibinfo {volume} {104}},\ \bibinfo {pages} {123024} (\bibinfo {year} {2021})},\ \Eprint {https://arxiv.org/abs/2110.11171} {arXiv:2110.11171 [gr-qc]} \BibitemShut {NoStop}%
\bibitem [{\citenamefont {{LIGO Scientific Collaboration and Virgo Collaboration}}(2017{\natexlab{a}})}]{LVC_GCN_NoticeG298048}%
  \BibitemOpen
  \bibfield  {author} {\bibinfo {author} {\bibnamefont {{LIGO Scientific Collaboration and Virgo Collaboration}}},\ }\href@noop {} {\bibinfo {title} {{GCN/LVC Notice: G298048 (LVC Initial Skymap)}}},\ \bibinfo {howpublished} {{General Coordinates Network (GCN/LVC) Notice}} (\bibinfo {year} {2017}{\natexlab{a}}),\ \bibinfo {note} {\url{https://gcn.gsfc.nasa.gov/notices_l/G298048.lvc}}\BibitemShut {NoStop}%
\bibitem [{\citenamefont {Essick}\ and\ \citenamefont {{LIGO Scientific Collaboration and Virgo Collaboration}}(2017)}]{LVC_GCN_Circ21505}%
  \BibitemOpen
  \bibfield  {author} {\bibinfo {author} {\bibfnamefont {R.~C.}\ \bibnamefont {Essick}}\ and\ \bibinfo {author} {\bibnamefont {{LIGO Scientific Collaboration and Virgo Collaboration}}},\ }\href@noop {} {\bibinfo {title} {{LIGO/Virgo G298048: Fermi GBM trigger 524666471/170817529: LIGO/Virgo Identification of a possible gravitational-wave counterpart}}},\ \bibinfo {howpublished} {{GCN Circular 21505}} (\bibinfo {year} {2017}),\ \bibinfo {note} {\url{https://gcn.nasa.gov/circulars/21505}}\BibitemShut {NoStop}%
\bibitem [{\citenamefont {{IGWN Alert Infrastructure Team}}(2023)}]{IGWN_UserGuide_Alerts}%
  \BibitemOpen
  \bibfield  {author} {\bibinfo {author} {\bibnamefont {{IGWN Alert Infrastructure Team}}},\ }\href@noop {} {\bibinfo {title} {{IGWN Public Alerts User Guide: Alert Contents \& Early-Warning Notices}}},\ \bibinfo {howpublished} {{Online documentation}} (\bibinfo {year} {2023}),\ \bibinfo {note} {\url{https://emfollow.docs.ligo.org/userguide/content.html}; \url{https://rtd.igwn.org/projects/userguide/en/latest/early_warning.html}}\BibitemShut {NoStop}%

\bibitem [{\citenamefont {{LIGO Scientific Collaboration and Virgo Collaboration}}(2017{\natexlab{b}})}]{45h2017}%
  \BibitemOpen
  \bibfield  {author} {\bibinfo {author} {\bibnamefont {{LIGO Scientific Collaboration and Virgo Collaboration}}},\ }\href@noop {} {\bibinfo {title} {Further analysis of a binary neutron star candidate with updated sky localization}},\ \bibinfo {howpublished} {GCN Circular} (\bibinfo {year} {2017}{\natexlab{b}}),\ \bibinfo {note} {available at \url{https://gcn.gsfc.nasa.gov/gcn3/21513.gcn3}}\BibitemShut {NoStop}%
\bibitem [{\citenamefont {Sachdev}\ \emph {et~al.}(2020)\citenamefont {Sachdev} \emph {et~al.}}]{Sachdeve2020}%
  \BibitemOpen
  \bibfield  {author} {\bibinfo {author} {\bibfnamefont {S.}~\bibnamefont {Sachdev}} \emph {et~al.},\ }\bibfield  {title} {\bibinfo {title} {{An Early-warning System for Electromagnetic Follow-up of Gravitational-wave Events}},\ }\href {https://doi.org/10.3847/2041-8213/abc753} {\bibfield  {journal} {\bibinfo  {journal} {Astrophys. J. Lett.}\ }\textbf {\bibinfo {volume} {905}},\ \bibinfo {pages} {L25} (\bibinfo {year} {2020})},\ \Eprint {https://arxiv.org/abs/2008.04288} {arXiv:2008.04288 [astro-ph.HE]} \BibitemShut {NoStop}%
  
  \bibitem [{\citenamefont {Pankow}\ \emph {et~al.}(2018)\citenamefont {Pankow} \emph {et~al.}}]{Pankow:2018qpo}%
  \BibitemOpen
  \bibfield  {author} {\bibinfo {author} {\bibfnamefont {C.}~\bibnamefont {Pankow}} \emph {et~al.},\ }\bibfield  {title} {\bibinfo {title} {{Mitigation of the instrumental noise transient in gravitational-wave data surrounding GW170817}},\ }\href {https://doi.org/10.1103/PhysRevD.98.084016} {\bibfield  {journal} {\bibinfo  {journal} {Phys. Rev. D}\ }\textbf {\bibinfo {volume} {98}},\ \bibinfo {pages} {084016} (\bibinfo {year} {2018})},\ \Eprint {https://arxiv.org/abs/1808.03619} {arXiv:1808.03619 [gr-qc]} \BibitemShut {NoStop}%
  
\bibitem [{\citenamefont {Singer}\ and\ \citenamefont {Price}(2016)}]{2016PhRvD..93b4013S}%
  \BibitemOpen
  \bibfield  {author} {\bibinfo {author} {\bibfnamefont {L.~P.}\ \bibnamefont {Singer}}\ and\ \bibinfo {author} {\bibfnamefont {L.~R.}\ \bibnamefont {Price}},\ }\bibfield  {title} {\bibinfo {title} {{Rapid Bayesian position reconstruction for gravitational-wave transients}},\ }\href {https://doi.org/10.1103/PhysRevD.93.024013} {\bibfield  {journal} {\bibinfo  {journal} {Phys. Rev. D}\ }\textbf {\bibinfo {volume} {93}},\ \bibinfo {pages} {024013} (\bibinfo {year} {2016})},\ \Eprint {https://arxiv.org/abs/1508.03634} {arXiv:1508.03634 [gr-qc]} \BibitemShut {NoStop}%
  
  \bibitem [{\citenamefont {Kawaguchi}\ \emph {et~al.}(2016)\citenamefont {Kawaguchi}, \citenamefont {Kyutoku}, \citenamefont {Shibata},\ and\ \citenamefont {Tanaka}}]{Kawaguchi2016_ApJ}%
  \BibitemOpen
  \bibfield  {author} {\bibinfo {author} {\bibfnamefont {K.}~\bibnamefont {Kawaguchi}}, \bibinfo {author} {\bibfnamefont {K.}~\bibnamefont {Kyutoku}}, \bibinfo {author} {\bibfnamefont {M.}~\bibnamefont {Shibata}},\ and\ \bibinfo {author} {\bibfnamefont {M.}~\bibnamefont {Tanaka}},\ }\bibfield  {title} {\bibinfo {title} {{Models of Kilonova/macronova Emission from Black Hole--Neutron Star Mergers}},\ }\href {https://doi.org/10.3847/0004-637X/825/1/52} {\bibfield  {journal} {\bibinfo  {journal} {Astrophys. J.}\ }\textbf {\bibinfo {volume} {825}},\ \bibinfo {pages} {52} (\bibinfo {year} {2016})},\ \Eprint {https://arxiv.org/abs/1601.07711} {arXiv:1601.07711 [astro-ph.HE]} \BibitemShut {NoStop}%
 
\bibitem [{\citenamefont {Shvartzvald}\ \emph {et~al.}(2024)\citenamefont {Shvartzvald} \emph {et~al.}}]{ultrasat2025}%
  \BibitemOpen
  \bibfield  {author} {\bibinfo {author} {\bibfnamefont {Y.}~\bibnamefont {Shvartzvald}} \emph {et~al.},\ }\bibfield  {title} {\bibinfo {title} {{ULTRASAT: A Wide-field Time-domain UV Space Telescope}},\ }\href {https://doi.org/10.3847/1538-4357/ad2704} {\bibfield  {journal} {\bibinfo  {journal} {Astrophys. J.}\ }\textbf {\bibinfo {volume} {964}},\ \bibinfo {pages} {74} (\bibinfo {year} {2024})},\ \Eprint {https://arxiv.org/abs/2304.14482} {arXiv:2304.14482 [astro-ph.IM]} \BibitemShut {NoStop}%
\bibitem [{\citenamefont {Song}\ \emph {et~al.}(2012)\citenamefont {Song} \emph {et~al.}}]{2012ApJ...761...22S}%
  \BibitemOpen
  \bibfield  {author} {\bibinfo {author} {\bibfnamefont {J.}~\bibnamefont {Song}} \emph {et~al.},\ }\bibfield  {title} {\bibinfo {title} {Redshifts, sample purity, and bcg positions for the galaxy cluster catalog from the first 720 square degrees of the south pole telescope survey},\ }\href {https://doi.org/10.1088/0004-637x/761/1/22} {\bibfield  {journal} {\bibinfo  {journal} {The Astrophysical Journal}\ }\textbf {\bibinfo {volume} {761}},\ \bibinfo {pages} {22} (\bibinfo {year} {2012})}\BibitemShut {NoStop}%

\bibitem [{\citenamefont {Ofek}\ \emph {et~al.}(2023)\citenamefont {Ofek} \emph {et~al.}}]{Ofek_2023}%
  \BibitemOpen
  \bibfield  {author} {\bibinfo {author} {\bibfnamefont {E.~O.}\ \bibnamefont {Ofek}} \emph {et~al.},\ }\bibfield  {title} {\bibinfo {title} {The large array survey telescope—system overview and performances},\ }\href {https://doi.org/10.1088/1538-3873/acd8f0} {\bibfield  {journal} {\bibinfo  {journal} {Publications of the Astronomical Society of the Pacific}\ }\textbf {\bibinfo {volume} {135}},\ \bibinfo {pages} {065001} (\bibinfo {year} {2023})}\BibitemShut {NoStop}%
\bibitem [{ras(2025)}]{rasa11v2}%
  \BibitemOpen
  \href@noop {} {\bibinfo {title} {{11'' Rowe-Ackermann Schmidt Astrograph (RASA 11) V2 Optical Tube Assembly (CGE Dovetail)}}} (\bibinfo {year} {2025}),\ \bibinfo {note} {\url{https://www.celestron.com/products/11-rowe-ackermann-schmidt-astrograph-rasa-11-v2-optical-tube-assembly-cge-dovetail}}\BibitemShut {NoStop}%
\bibitem [{\citenamefont {Abbott}\ \emph {et~al.}(2009)\citenamefont {Abbott} \emph {et~al.}}]{Abbott2009}%
  \BibitemOpen
  \bibfield  {author} {\bibinfo {author} {\bibfnamefont {B.~P.}\ \bibnamefont {Abbott}} \emph {et~al.},\ }\bibfield  {title} {\bibinfo {title} {Rep. prog. phys. 72 076901},\ }\href {https://doi.org/10.1088/0034-4885/72/7/076901} {\bibfield  {journal} {\bibinfo  {journal} {Rep. Prog. Phys.}\ }\textbf {\bibinfo {volume} {72}},\ \bibinfo {pages} {076901} (\bibinfo {year} {2009})},\ \bibinfo {note} {available at \url{https://iopscience.iop.org/article/10.1088/0034-4885/72/7/076901}}\BibitemShut {NoStop}%
\bibitem [{\citenamefont {Caron}\ \emph {et~al.}(1997)\citenamefont {Caron} \emph {et~al.}}]{Caron1997}%
  \BibitemOpen
  \bibfield  {author} {\bibinfo {author} {\bibfnamefont {B.}~\bibnamefont {Caron}} \emph {et~al.},\ }\bibfield  {title} {\bibinfo {title} {Class. quantum grav. 14 146},\ }\href {https://doi.org/10.1088/0264-9381/14/6/011} {\bibfield  {journal} {\bibinfo  {journal} {Class. Quantum Grav.}\ }\textbf {\bibinfo {volume} {14}},\ \bibinfo {pages} {146} (\bibinfo {year} {1997})},\ \bibinfo {note} {available at \url{https://iopscience.iop.org/article/10.1088/0264-9381/14/6/011/pdf}}\BibitemShut {NoStop}%
\bibitem [{\citenamefont {Akutsu}\ \emph {et~al.}(2019)\citenamefont {Akutsu} \emph {et~al.}}]{KAGRA2019}%
  \BibitemOpen
  \bibfield  {author} {\bibinfo {author} {\bibfnamefont {T.}~\bibnamefont {Akutsu}} \emph {et~al.} (\bibinfo {collaboration} {KAGRA}),\ }\bibfield  {title} {\bibinfo {title} {{KAGRA: 2.5 Generation Interferometric Gravitational Wave Detector}},\ }\href {https://doi.org/10.1038/s41550-018-0658-y} {\bibfield  {journal} {\bibinfo  {journal} {Nature Astron.}\ }\textbf {\bibinfo {volume} {3}},\ \bibinfo {pages} {35} (\bibinfo {year} {2019})},\ \Eprint {https://arxiv.org/abs/1811.08079} {arXiv:1811.08079 [gr-qc]} \BibitemShut {NoStop}%
\bibitem [{\citenamefont {Chaudhary}\ \emph {et~al.}(2024)\citenamefont
 {Chaudhary} \emph {et~al.}}]{Chaudhary:2023vec}%
  \BibitemOpen
  \bibfield  {author} {\bibinfo {author} {\bibfnamefont {S.~S.}~\bibnamefont {Chaudhary}} \emph {et~al.},\ }\bibfield  {title} {\bibinfo {title} {{Low-latency gravitational wave alert products and their performance at the time of the fourth LIGO-Virgo-KAGRA observing run}},\ }\href {https://doi.org/10.1073/pnas.2316474121} {\bibfield  {journal} {\bibinfo  {journal} {Proc. Nat. Acad. Sci.}\ }\textbf {\bibinfo {volume} {121}},\ \bibinfo {pages} {e2316474121} (\bibinfo {year} {2024})},\ \Eprint {https://arxiv.org/abs/2308.04545} {arXiv:2308.04545 [astro-ph.HE]} \BibitemShut {NoStop}%

  \bibitem [{\citenamefont {Christensen}\ and\ \citenamefont {Meyer}(2022)}]{Christensen:2022bxb}%
  \BibitemOpen
  \bibfield  {author} {\bibinfo {author} {\bibfnamefont {N.}~\bibnamefont {Christensen}}\ and\ \bibinfo {author} {\bibfnamefont {R.}~\bibnamefont {Meyer}},\ }\bibfield  {title} {\bibinfo {title} {{Parameter estimation with gravitational waves}},\ }\href {https://doi.org/10.1103/RevModPhys.94.025001} {\bibfield  {journal} {\bibinfo  {journal} {Rev. Mod. Phys.}\ }\textbf {\bibinfo {volume} {94}},\ \bibinfo {pages} {025001} (\bibinfo {year} {2022})},\ \Eprint {https://arxiv.org/abs/2204.04449} {arXiv:2204.04449 [gr-qc]} \BibitemShut {NoStop}%
 
\bibitem [{\citenamefont {Duverne}\ \emph {et~al.}(2024)\citenamefont {Duverne} \emph {et~al.}}]{Duverne:2023joq}%
  \BibitemOpen
  \bibfield  {author} {\bibinfo {author} {\bibfnamefont {P.-A.}\ \bibnamefont {Duverne}} \emph {et~al.},\ }\bibfield  {title} {\bibinfo {title} {{Optimizing the low-latency localization of gravitational waves}},\ }\href {https://doi.org/10.1103/PhysRevD.110.102002} {\bibfield  {journal} {\bibinfo  {journal} {Phys. Rev. D}\ }\textbf {\bibinfo {volume} {110}},\ \bibinfo {pages} {102002} (\bibinfo {year} {2024})},\ \Eprint {https://arxiv.org/abs/2312.15457} {arXiv:2312.15457 [gr-qc]} \BibitemShut {NoStop}%

\bibitem [{\citenamefont {Coughlin}\ \emph {et~al.}(2018)\citenamefont
  {Coughlin} \emph {et~al.}}]{Coughlin:2018lta}%
  \BibitemOpen
  \bibfield  {author} {\bibinfo {author} {\bibfnamefont {M.~W.}\ \bibnamefont
  {Coughlin}} \emph {et~al.},\ }\bibfield  {title} {\bibinfo {title}
  {{Optimizing searches for electromagnetic counterparts of gravitational wave
  triggers}},\ }\href {https://doi.org/10.1093/mnras/sty1066} {\bibfield
  {journal} {\bibinfo  {journal} {Mon. Not. R. Astron. Soc.}\ }\textbf
  {\bibinfo {volume} {478}},\ \bibinfo {pages} {692} (\bibinfo {year}
  {2018})},\ \Eprint {https://arxiv.org/abs/1803.02255} {arXiv:1803.02255
  [astro-ph.IM]} \BibitemShut {NoStop}%

\bibitem [{\citenamefont {Coughlin}\ \emph {et~al.}(2019)\citenamefont
  {Coughlin} \emph {et~al.}}]{Coughlin:2019qkn}%
  \BibitemOpen
  \bibfield  {author} {\bibinfo {author} {\bibfnamefont {M.~W.}\ \bibnamefont
  {Coughlin}} \emph {et~al.},\ }\bibfield  {title} {\bibinfo {title}
  {{Optimizing multitelescope observations of gravitational-wave
  counterparts}},\ }\href {https://doi.org/10.1093/mnras/stz2485} {\bibfield
  {journal} {\bibinfo  {journal} {Mon. Not. R. Astron. Soc.}\ }\textbf
  {\bibinfo {volume} {489}},\ \bibinfo {pages} {5775} (\bibinfo {year}
  {2019})},\ \Eprint {https://arxiv.org/abs/1909.01244} {arXiv:1909.01244
  [astro-ph.IM]} \BibitemShut {NoStop}%

\bibitem [{\citenamefont {Almualla}\ \emph {et~al.}(2020)\citenamefont
  {Almualla} \emph {et~al.}}]{Almualla:2020hbs}%
  \BibitemOpen
  \bibfield  {author} {\bibinfo {author} {\bibfnamefont {M.}~\bibnamefont
  {Almualla}} \emph {et~al.},\ }\bibfield  {title} {\bibinfo {title} {{Dynamic
  scheduling: Target of opportunity observations of gravitational wave
  events}},\ }\href {https://doi.org/10.1093/mnras/staa1264} {\bibfield
  {journal} {\bibinfo  {journal} {Mon. Not. R. Astron. Soc.}\ }\textbf
  {\bibinfo {volume} {495}},\ \bibinfo {pages} {4366} (\bibinfo {year}
  {2020})},\ \Eprint {https://arxiv.org/abs/2003.09718} {arXiv:2003.09718
  [astro-ph.IM]} \BibitemShut {NoStop}%

\bibitem [{\citenamefont {Seglar-Arroyo}\ \emph {et~al.}(2024)\citenamefont
  {Seglar-Arroyo} \emph {et~al.}}]{SeglarArroyo:2024tilepy}%
  \BibitemOpen
  \bibfield  {author} {\bibinfo {author} {\bibfnamefont {M.}~\bibnamefont
  {Seglar-Arroyo}} \emph {et~al.},\ }\bibfield  {title} {\bibinfo {title}
  {{Cross-Observatory Coordination with tilepy: A Novel Tool for Observations
  of Multi-Messenger Transient Events}},\ }\href
  {https://doi.org/10.3847/1538-4365/ad5bde} {\bibfield  {journal} {\bibinfo
  {journal} {Astrophys. J. Suppl. Ser.}\ }\textbf {\bibinfo {volume} {274}},\
  \bibinfo {pages} {1} (\bibinfo {year} {2024})},\ \Eprint
  {https://arxiv.org/abs/2407.18076} {arXiv:2407.18076 [astro-ph.IM]}
  \BibitemShut {NoStop}%

\bibitem [{\citenamefont {Singer}\ \emph {et~al.}(2025)\citenamefont {Singer}
  \emph {et~al.}}]{Singer:2025UVEX}%
  \BibitemOpen
  \bibfield  {author} {\bibinfo {author} {\bibfnamefont {L.~P.}\ \bibnamefont
  {Singer}} \emph {et~al.},\ }\bibfield  {title} {\bibinfo {title} {{Optimal
  follow-up of gravitational-wave events with the UltraViolet EXplorer
  (UVEX)}},\ }\href {https://doi.org/10.1088/1538-3873/adcfc6} {\bibfield
  {journal} {\bibinfo  {journal} {Publ. Astron. Soc. Pac.}\ }\textbf {\bibinfo
  {volume} {137}},\ \bibinfo {pages} {074501} (\bibinfo {year} {2025})},\
  \Eprint {https://arxiv.org/abs/2502.17560} {arXiv:2502.17560 [astro-ph.IM]}
  \BibitemShut {NoStop}%

\bibitem [{\citenamefont {Sachdev}\ \emph {et~al.}(2020)\citenamefont
  {Sachdev} \emph {et~al.}}]{Sachdev:2020lfd}%
  \BibitemOpen
  \bibfield  {author} {\bibinfo {author} {\bibfnamefont {S.}~\bibnamefont
  {Sachdev}} \emph {et~al.},\ }\bibfield  {title} {\bibinfo {title} {{An
  early-warning system for electromagnetic follow-up of gravitational-wave
  events}},\ }\href {https://doi.org/10.3847/2041-8213/abc753} {\bibfield
  {journal} {\bibinfo  {journal} {Astrophys. J. Lett.}\ }\textbf {\bibinfo
  {volume} {905}},\ \bibinfo {pages} {L25} (\bibinfo {year} {2020})},\ \Eprint
  {https://arxiv.org/abs/2008.04288} {arXiv:2008.04288 [astro-ph.HE]}
  \BibitemShut {NoStop}%

\bibitem [{\citenamefont {Magee}\ \emph {et~al.}(2021)\citenamefont {Magee}
  \emph {et~al.}}]{Magee:2021EW}%
  \BibitemOpen
  \bibfield  {author} {\bibinfo {author} {\bibfnamefont {R.}~\bibnamefont
  {Magee}} \emph {et~al.},\ }\bibfield  {title} {\bibinfo {title} {{First
  demonstration of early warning gravitational-wave alerts}},\ }\href
  {https://doi.org/10.3847/2041-8213/abed54} {\bibfield  {journal} {\bibinfo
  {journal} {Astrophys. J. Lett.}\ }\textbf {\bibinfo {volume} {910}},\
  \bibinfo {pages} {L21} (\bibinfo {year} {2021})},\ \Eprint
  {https://arxiv.org/abs/2102.04555} {arXiv:2102.04555 [astro-ph.HE]}
  \BibitemShut {NoStop}%

\bibitem [{\citenamefont {Nitz}\ \emph {et~al.}(2020)\citenamefont {Nitz},
  \citenamefont {Schäfer},\ and\ \citenamefont
  {Dal~Canton}}]{Nitz:2020abbc10}%
  \BibitemOpen
  \bibfield  {author} {\bibinfo {author} {\bibfnamefont {A.~H.}\ \bibnamefont
  {Nitz}}, \bibinfo {author} {\bibfnamefont {M.}~\bibnamefont {Schäfer}},\
  and\ \bibinfo {author} {\bibfnamefont {T.}~\bibnamefont {Dal~Canton}},\
  }\bibfield  {title} {\bibinfo {title} {{Gravitational-wave merger
  forecasting: Scenarios for the early detection and localization of
  compact-binary mergers with ground-based observatories}},\ }\href
  {https://doi.org/10.3847/2041-8213/abbc10} {\bibfield  {journal} {\bibinfo
  {journal} {Astrophys. J. Lett.}\ }\textbf {\bibinfo {volume} {902}},\
  \bibinfo {pages} {L29} (\bibinfo {year} {2020})},\ \Eprint
  {https://arxiv.org/abs/2009.04439} {arXiv:2009.04439 [astro-ph.HE]}
  \BibitemShut {NoStop}%
\bibitem [{\citenamefont {Petrov}\ \emph {et~al.}(2022)\citenamefont {Petrov}
  \emph {et~al.}}]{Petrov:2021bqm}%
  \BibitemOpen
  \bibfield  {author} {\bibinfo {author} {\bibfnamefont {P.}~\bibnamefont
  {Petrov}} \emph {et~al.},\ }\bibfield  {title} {\bibinfo {title} {{Data-driven
  expectations for electromagnetic counterpart searches based on LIGO/Virgo
  public alerts}},\ }\href {https://doi.org/10.3847/1538-4357/ac366d}
  {\bibfield  {journal} {\bibinfo  {journal} {Astrophys. J.}\ }\textbf
  {\bibinfo {volume} {924}},\ \bibinfo {pages} {54} (\bibinfo {year}
  {2022})},\ \Eprint {https://arxiv.org/abs/2108.07277} {arXiv:2108.07277
  [astro-ph.HE]} \BibitemShut {NoStop}%

\bibitem [{\citenamefont {Kiendrebeogo}\ \emph {et~al.}(2023)\citenamefont
  {Kiendrebeogo} \emph {et~al.}}]{Kiendrebeogo:2023hzf}%
  \BibitemOpen
  \bibfield  {author} {\bibinfo {author} {\bibfnamefont {R.~W.}\ \bibnamefont
  {Kiendrebeogo}} \emph {et~al.},\ }\bibfield  {title} {\bibinfo {title}
  {{Updated observing scenarios and multimessenger implications for the
  international gravitational-wave networks O4 and O5}},\ }\href
  {https://doi.org/10.3847/1538-4357/acfcb1} {\bibfield  {journal} {\bibinfo
  {journal} {Astrophys. J.}\ }\textbf {\bibinfo {volume} {958}},\ \bibinfo
  {pages} {158} (\bibinfo {year} {2023})},\ \Eprint
  {https://arxiv.org/abs/2306.09234} {arXiv:2306.09234 [astro-ph.HE]}
  \BibitemShut {NoStop}%
\bibitem [{\citenamefont {Abbott}\ \emph {et~al.}(2016)\citenamefont
  {Abbott} \emph {et~al.}}]{KAGRA:2013rdx}%
  \BibitemOpen
  \bibfield  {author} {\bibinfo {author} {\bibfnamefont {B.~P.}\ \bibnamefont
  {Abbott}} \emph {et~al.} (KAGRA, LIGO Scientific, Virgo),\ }\bibfield  {title}
  {\bibinfo {title} {{Prospects for observing and localizing gravitational-wave
  transients with Advanced LIGO, Advanced Virgo and KAGRA}},\ }\href
  {https://doi.org/10.1007/s41114-020-00026-9} {\bibfield  {journal} {\bibinfo
  {journal} {Living Rev. Rel.}\ }\textbf {\bibinfo {volume} {19}},\ \bibinfo
  {pages} {1} (\bibinfo {year} {2016})},\ \Eprint
  {https://arxiv.org/abs/1304.0670} {arXiv:1304.0670 [gr-qc]}%
  \BibitemShut {NoStop}%

\bibitem [{\citenamefont {Abbott}\ \emph {et~al.}(2017{\natexlab{f}})\citenamefont {Abbott} \emph {et~al.}}]{2017AnP...52900209A}%
  \BibitemOpen
  \bibfield  {author} {\bibinfo {author} {\bibfnamefont {B.~P.}\ \bibnamefont {Abbott}} \emph {et~al.} (\bibinfo {collaboration} {LIGO Scientific, Virgo}),\ }\bibfield  {title} {\bibinfo {title} {{The basic physics of the binary black hole merger GW150914}},\ }\href {https://doi.org/10.1002/andp.201600209} {\bibfield  {journal} {\bibinfo  {journal} {Annalen Phys.}\ }\textbf {\bibinfo {volume} {529}},\ \bibinfo {pages} {1600209} (\bibinfo {year} {2017}{\natexlab{f}})},\ \Eprint {https://arxiv.org/abs/1608.01940} {arXiv:1608.01940 [gr-qc]} \BibitemShut {NoStop}%
\bibitem [{\citenamefont {Kacanja}\ and\ \citenamefont {Nitz}(2025)}]{arxiv:2412.05369v1}%
  \BibitemOpen
  \bibfield  {author} {\bibinfo {author} {\bibfnamefont {K.}~\bibnamefont {Kacanja}}\ and\ \bibinfo {author} {\bibfnamefont {A.~H.}\ \bibnamefont {Nitz}},\ }\href {https://doi.org/10.3847/1538-4357/adc9a5} {\bibinfo {title} {{A Search for Low-mass Neutron Stars in the Third Observing Run of Advanced LIGO and Virgo}}} (\bibinfo {year} {2025}),\ \Eprint {https://arxiv.org/abs/2412.05369} {arXiv:2412.05369 [astro-ph.HE]} \BibitemShut {NoStop}%
\bibitem [{\citenamefont {{Antelis}}\ and\ \citenamefont {{Moreno}}(2017)}]{2017EPJP..132...10A}%
  \BibitemOpen
  \bibfield  {author} {\bibinfo {author} {\bibfnamefont {J.~M.}\ \bibnamefont {{Antelis}}}\ and\ \bibinfo {author} {\bibfnamefont {C.}~\bibnamefont {{Moreno}}},\ }\bibfield  {title} {\bibinfo {title} {{Obtaining gravitational waves from inspiral binary systems using LIGO data}},\ }\href {https://doi.org/10.1140/epjp/i2017-11283-5} {\bibfield  {journal} {\bibinfo  {journal} {European Physical Journal Plus}\ }\textbf {\bibinfo {volume} {132}},\ \bibinfo {eid} {10} (\bibinfo {year} {2017})}\BibitemShut {NoStop}%
\bibitem [{\citenamefont {Bernuzzi}\ \emph {et~al.}(2012)\citenamefont {Bernuzzi}, \citenamefont {Thierfelder},\ and\ \citenamefont {Brügmann}}]{Bernuzzi_2012}%
  \BibitemOpen
  \bibfield  {author} {\bibinfo {author} {\bibfnamefont {S.}~\bibnamefont {Bernuzzi}}, \bibinfo {author} {\bibfnamefont {M.}~\bibnamefont {Thierfelder}},\ and\ \bibinfo {author} {\bibfnamefont {B.}~\bibnamefont {Brügmann}},\ }\bibfield  {title} {\bibinfo {title} {Accuracy of numerical relativity waveforms from binary neutron star mergers and their comparison with post-newtonian waveforms},\ }\bibfield  {journal} {\bibinfo  {journal} {Physical Review D}\ }\textbf {\bibinfo {volume} {85}},\ \href {https://doi.org/10.1103/physrevd.85.104030} {10.1103/physrevd.85.104030} (\bibinfo {year} {2012})\BibitemShut {NoStop}%
\bibitem [{\citenamefont {Baiotti}\ \emph {et~al.}(2011)\citenamefont {Baiotti}, \citenamefont {Damour}, \citenamefont {Giacomazzo}, \citenamefont {Nagar},\ and\ \citenamefont {Rezzolla}}]{Baiotti_2011}%
  \BibitemOpen
  \bibfield  {author} {\bibinfo {author} {\bibfnamefont {L.}~\bibnamefont {Baiotti}}, \bibinfo {author} {\bibfnamefont {T.}~\bibnamefont {Damour}}, \bibinfo {author} {\bibfnamefont {B.}~\bibnamefont {Giacomazzo}}, \bibinfo {author} {\bibfnamefont {A.}~\bibnamefont {Nagar}},\ and\ \bibinfo {author} {\bibfnamefont {L.}~\bibnamefont {Rezzolla}},\ }\bibfield  {title} {\bibinfo {title} {Accurate numerical simulations of inspiralling binary neutron stars and their comparison with effective-one-body analytical models},\ }\bibfield  {journal} {\bibinfo  {journal} {Physical Review D}\ }\textbf {\bibinfo {volume} {84}},\ \href {https://doi.org/10.1103/physrevd.84.024017} {10.1103/physrevd.84.024017} (\bibinfo {year} {2011})\BibitemShut {NoStop}%
\bibitem [{\citenamefont {{LIGO Scientific Collaboration}}(2024)}]{LALSuite}%
  \BibitemOpen
  \bibfield  {author} {\bibinfo {author} {\bibnamefont {{LIGO Scientific Collaboration}}},\ } {\bibinfo {title} {Lalsuite: Lalinspiral documentation}},\ \bibinfo {howpublished} {\url{https://lscsoft.docs.ligo.org/lalsuite/lalinspiral/group__lalinspiral__inspiral.html}} (\bibinfo {year} {2024})\BibitemShut {NoStop}%
\bibitem [{\citenamefont {Barsotti}\ \emph {et~al.}(2018)\citenamefont {Barsotti} \emph {et~al.}}]{LIGO_T1800044}%
  \BibitemOpen
  \bibfield  {author} {\bibinfo {author} {\bibfnamefont {L.}~\bibnamefont {Barsotti}} \emph {et~al.},\ }\href {https://dcc.ligo.org/public/0149/T1800044/005/T1800044-v5.pdf} {\bibinfo {title} {Ligo document t1800044-v5}} (\bibinfo {year} {2018})\BibitemShut {NoStop}%
  \bibitem [{\citenamefont {Di~Cesare}(2025)}]{DiCesare:2025wnb}%
  \BibitemOpen
  \bibfield  {author} {\bibinfo {author} {\bibfnamefont {M.}~\bibnamefont {Di~Cesare}},\ }%
  \bibfield  {title} {\bibinfo {title} {{Status of the O4 run and latest non-CBC results}},\ }%
  \href {} {\bibfield  {journal} {\bibinfo  {journal} {arXiv preprint}\ }}%
  (\bibinfo {year} {2025}),\ \Eprint {https://arxiv.org/abs/2505.18802} {arXiv:2505.18802 [gr-qc]}%
  \BibitemShut {NoStop}%

\bibitem [{\citenamefont {Collaboration}(2020)}]{LIGO_T2000012}%
  \BibitemOpen
  \bibfield  {author} {\bibinfo {author} {\bibfnamefont {L.~S.}\ \bibnamefont {Collaboration}},\ }\href {https://dcc.ligo.org/LIGO-T2000012/public} {\bibinfo {title} {Ligo document t2000012}} (\bibinfo {year} {2020})\BibitemShut {NoStop}%
\bibitem [{\citenamefont {Coulter}\ \emph {et~al.}(2017)\citenamefont
  {Coulter} \emph {et~al.}}]{Coulter:2017wya}%
  \BibitemOpen
  \bibfield  {author} {\bibinfo {author} {\bibfnamefont {D.~A.}\ \bibnamefont
  {Coulter}} \emph {et~al.},\ }\bibfield  {title} {\bibinfo {title} {{Swope
  Supernova Survey 2017a (SSS17a), the Optical Counterpart to a Gravitational
  Wave Source}},\ }\href {https://doi.org/10.1126/science.aap9811} {\bibfield
  {journal} {\bibinfo  {journal} {Science}\ }\textbf {\bibinfo {volume}
  {358}},\ \bibinfo {pages} {1556} (\bibinfo {year} {2017})},\ \Eprint
  {https://arxiv.org/abs/1710.05452} {arXiv:1710.05452 [astro-ph.HE]}
  \BibitemShut {NoStop}%

\bibitem [{\citenamefont {Israeli~Aerospace~Industry~(IAI)}\ and\
\citenamefont {Weizmann~Institute~of~Science}(2022)}]{ULTRASAT_IAI_2022}%
  \BibitemOpen
  \bibfield  {author} {\bibinfo {author} {\bibnamefont
  {Israeli~Aerospace~Industry~(IAI)}}\ and\ \bibinfo {author} {\bibnamefont
  {Weizmann~Institute~of~Science}},\ }\bibfield  {title} {\bibinfo {title}
  {{ULTRASAT System Design Overview}},\ }
  {\bibinfo {howpublished} {Online, ULTRASAT Project and Technology Overview
  (Accessed October 2025)}}\BibitemShut {NoStop}%
\bibitem [{\citenamefont {{Bellm}}\ \emph {et~al.}(2019)\citenamefont {{Bellm}} \emph {et~al.}}]{ZTF}%
  \BibitemOpen
  \bibfield  {author} {\bibinfo {author} {\bibfnamefont {E.~C.}\ \bibnamefont {{Bellm}}} \emph {et~al.},\ }\href {https://doi.org/10.1088/1538-3873/aaecbe} {\bibinfo {title} {{The Zwicky Transient Facility: System Overview, Performance, and First Results}}} (\bibinfo {year} {2019}),\ \Eprint {https://arxiv.org/abs/1902.01932} {arXiv:1902.01932 [astro-ph.IM]} \BibitemShut {NoStop}%
\bibitem [{\citenamefont {Collaboration}(2024{\natexlab{a}})}]{LIGO_Observing_Plan}%
  \BibitemOpen
  \bibfield  {author} {\bibinfo {author} {\bibfnamefont {L.~S.}\ \bibnamefont {Collaboration}},\ }\href {https://observing.docs.ligo.org/plan/} {\bibinfo {title} {Ligo observing plan}} (\bibinfo {year} {2024}{\natexlab{a}})\BibitemShut {NoStop}%
\bibitem [{\citenamefont {Collaboration}(2024{\natexlab{b}})}]{LIGO_early_warning}%
  \BibitemOpen
  \bibfield  {author} {\bibinfo {author} {\bibfnamefont {L.~S.}\ \bibnamefont {Collaboration}},\ }\href {https://emfollow.docs.ligo.org/userguide/early_warning.html} {\bibinfo {title} {Early warning user guide}} (\bibinfo {year} {2024}{\natexlab{b}})\BibitemShut {NoStop}%
\bibitem [{\citenamefont {Observatory}(2024)}]{LSST}%
  \BibitemOpen
  \bibfield  {author} {\bibinfo {author} {\bibfnamefont {V.~C.~R.}\ \bibnamefont {Observatory}},\ }\href {https://kipac.stanford.edu/research/projects/vera-rubin-observatorys-legacy-survey-space-and-time} {\bibinfo {title} {Vera c. rubin observatory's legacy survey of space and time}} (\bibinfo {year} {2024})\BibitemShut {NoStop}%

\end{thebibliography}

%

\end{document}